\documentclass[prd,nofootinbib,amsfonts,notitlepage]{revtex4-1}
\usepackage[utf8]{inputenc}
\usepackage[T1]{fontenc}
\usepackage{hyperref}
\usepackage{graphicx,bm,amsmath,color,scalerel}
\usepackage{bbold}
\usepackage{wasysym}
\usepackage{verbatim}
\newcommand{\be}{\begin{equation}}
\newcommand{\ee}{\end{equation}}
\newcommand{\beq}{\begin{eqnarray}}
\newcommand{\eeq}{\end{eqnarray}}
\newcommand{\ba}{\begin{align}}
\newcommand{\ea}{\end{align}}

\newcommand{\bigo}{\,\scaleobj{1.3}{\oplus}\,} %para hacer el \oplus grande, pero no tanto como el \bigoplus; el tamaño puede ajustarse. Hay que incluir en la definición los espacios que LaTeX pone automáticamente a los operadores, pero que no pone al definir este objeto nuevo
\newcommand{\ohat}{\,\hat{\oplus}\,}
\newcommand{\tildebigo}{\,\tilde{\scaleobj{1.3}{\oplus}}\,}

\newcommand{\p}{|p\rangle}
\newcommand{\pp}{|p'\rangle}
\newcommand{\pq}{|p;q\rangle}
\newcommand{\fj}{[f_j(p)]_\mu}
\newcommand{\pqp}{|p';q'\rangle}
\newcommand{\fja}{[f^{(1)}_j(p,q)]_\mu}
\newcommand{\fjb}{[f^{(2)}_j(p,q)]_\mu}
\newcommand{\cop}{(p\oplus q)_\mu}

\begin{document}

\title{Beyond Special Relativity at second order}

\author{J.M. Carmona}
\affiliation{Departamento de F\'{\i}sica Te\'orica,
Universidad de Zaragoza, Zaragoza 50009, Spain}
\author{J.L. Cort\'es}
\affiliation{Departamento de F\'{\i}sica Te\'orica,
Universidad de Zaragoza, Zaragoza 50009, Spain}
\author{J.J. Relancio}
\email{jcarmona@unizar.es, cortes@unizar.es, jjrelancio@gmail.com}
\affiliation{Departamento de F\'{\i}sica Te\'orica,
Universidad de Zaragoza, Zaragoza 50009, Spain}

\begin{abstract}
The study of generic, non-linear, deformations of Special Relativity parametrized by a high-energy scale $M$, which was carried out at first order in $M$ in Ref.~\cite{Carmona2012}, is extended to second order. This can be done systematically through a (`generalized') change of variables from momentum variables that transform linearly. We discuss the different perspectives on the meaning of the change of variables, obtain the coefficients of modified composition laws and Lorentz transformations at second order, and work out how $\kappa$-Poincaré, the most commonly used example in the literature, is reproduced as a particular case of the generic framework exposed here.
\end{abstract}

\maketitle

\section{Introduction}

The possibility to extend the usual Lorentz symmetry to a new relativistic invariance characterized by an observer-independent mass scale (or length scale) $M$ has been investigated for some time now in connection with a quantum theory of gravity, for which this mass (or length) scale would be the Planck mass, $m_P\approx 1.2\times 10^{19}$\,eV/$c^2$ (or Planck length, $\ell_P\approx 1.6\times 10^{-35}\,$m). This research has been mostly made in the context of the so-called DSR (for Doubly or Deformed Special Relativity) models (see Ref.~\cite{AmelinoCamelia:2010pd} for a review), initially formulated as relativistic theories with non-linear, $M$-dependent, deformations of the usual boosts of Special Relativity (SR) in momentum space. DSR was later interpreted geometrically as corresponding to a curved momentum space, which resulted in an observer-dependent notion of locality, named as the `principle of relative locality' \cite{AmelinoCamelia:2011bm,AmelinoCamelia:2011pe}. In fact, it was shown that the `quantum deformation' of the Poincar\'e group known as $\kappa$-Poincar\'e,\footnote{The deformation parameter, $\kappa$, plays the role of $1/M$.} an algebraic structure proposed by mathematicians years ago to describe the symmetries of a noncommutative space~\cite{Majid1994}, can be interpreted in the framework of curved momentum spaces leading to relative locality~\cite{Gubitosi:2013rna}. In particular, the co-product of the momentum and the co-product of boosts in $\kappa$-Poincar\'e (which can be seen as operators acting in the tensor product of two momentum spaces) have direct interpretations in the DSR language, in terms of the composition of momenta (which is necessarily non-linear, since a conservation law based on the linear addition of momenta would not be invariant under non-linear boosts) and the Lorentz transformations of momenta in a two-particle system (that, for a generic composition law of momenta, will be different from the Lorentz transformation in the one-particle sector), respectively.

There exists then a correspondence between the deformed Casimir of the algebra, the deformed Poincar\'e commutators, and the co-product structure as a Hopf algebra on one side, and the modified dispersion relation, modified (one-particle) Lorentz transformations, and the two-particle sector (boosts and composition law) of DSR on the other~\cite{Kowalski-Glikman2005}. However, the coassociativity axiom of Hopf algebras implies a restriction on the momentum composition rule, which must be associative~\cite{Gubitosi:2013rna}, so that one can say that the Hopf algebra framework is more restrictive than a general DSR or relative locality theory. 

The original examples of DSR theories~\cite{Amelino-Camelia2001,Amelino-Camelia2002a} as modifications of the transformation rules between different inertial observers were explored in leading order of the deformation (Planck-scale related) observer-independent scale. Later on, the connection between DSR and Hopf algebras referred above made possible to get all-order results~\cite{Bruno2001,Kowalski-Glikman2001} and several examples of `exact' DSR theories (named as DSR1, DSR2 or DSR3) were formulated~\cite{Amelino-Camelia2002,Magueijo2002,AmelinoCamelia:2002gv}. However, the general form of a DSR theory, as well as the classification of all the possible deformations of the Poincar\'e Hopf algebra is still an open problem~\cite{Bisio:2016jsw}.

A way to sort out all the possible DSR theories is the study of the mathematical relations that the relativity principle imposes among the coefficients of the deformed dispersion relation, the deformed Lorentz transformations, and the non-linear composition law. An exploratory study was initiated in Ref.~\cite{AmelinoCamelia:2011yi} at first order in the deformation scale, and then carried out systematically at the same order in Ref.~\cite{Carmona2012} (both type of studies were generalized, also at first order, to the case of non-universal kinematics, in Refs.~\cite{Amelino-Camelia2012}  and Ref.~\cite{Carmona:2014aba}, respectively). All these works restricted their studies to general deformations of Special Relativity at first order in $1/M$ compatible with standard rotational invariance. 

Ref.~\cite{Carmona2012} contained also a short discussion on the outcome of a change of variables in momentum space. In fact, since the initial development of DSR some authors have expressed their concern that DSR might be nothing but classical SR in a non-linear disguise, since one can always make a change of variables to new momentum variables that transform linearly~\cite{Lukierski:2002df,Ahluwalia2002a}. However, these remarks usually forget about the non-trivial multiparticle sector of DSR theories~\cite{Amelino-Camelia2002a,Amelino-Camelia2002}. From the algebraic point of view, a change of variables in momentum space corresponds to a change of basis, and indeed there exists the so-called `classical basis' of $\kappa$-Poincar\'e, in which the boost-momentum commutators, together with the Lorentz sector, form the classical Poincar\'e algebra; however, it is clear that $\kappa$-Poincar\'e is not equivalent to the standard Poincar\'e algebra, since the classical basis referred above has a non-trivial co-algebra structure~\cite{Kowalski-Glikman2002}. 

From the point of view of DSR, it is normally assumed that the momentum variables that transform non-linearly would be somehow distinguished by the dynamics as the `physical' momentum variables, and even different basis of $\kappa$-Poincar\'e (which are mathematically equivalent, although their physical equivalence is a subject of debate~\cite{Kowalski-Glikman2002}) would correspond to different physical theories. The DSR point of view of a change of variables is, therefore, not completely equivalent to the algebraic point of view.

Moreover, in Ref.~\cite{Carmona2012} it was shown that such a change of variables is able to reduce the (first-order) modified kinematics (including its multiparticle sector) to that of SR if and only if the momentum composition law is commutative. In the geometric interpretation of relative locality, a commutative composition law corresponds to a torsion-less momentum space~\cite{AmelinoCamelia:2011bm}, and a change of momentum variables is just a change of coordinates in momentum space, with no relevance in its geometric properties. Therefore, the previous result corresponds to saying that at first order in $1/M$ (when one is not sensitive to the curvature of momentum space, which enters only at $M^{-2}$,~\cite{AmelinoCamelia:2011bm}), the momentum space is Minkowski if the torsion is zero~\cite{Carmona2012}. From the point of view of DSR, deformations of SR at leading order with commutative composition laws are different from SR; from the algebraic (Hopf algebras) and geometric (relative locality) points of view they would be equivalent to SR, since in these frameworks the change of variables that reduce the theory to SR can be interpreted as a change of basis or a change of coordinates, respectively.\footnote{It should be noted that $\kappa$-Poincar\'e is an example of a Hopf algebra whose co-product of momentum (that is, the momentum composition law) is not commutative: that is why the classical basis of $\kappa$-Poincar\'e is not completely trivial, as mentioned before, and $\kappa$-Poincar\'e is not equivalent to SR, even at first order in $\kappa$.}

Independently of the adopted point of view (or which is the correct, physical, one), geometric, algebraic, or DSR, one could take the change of variables as a mere mathematical tool that simplifies the problem of finding the different theories which are compatible with a relativity principle, that is: a deformed theory that results from SR from a change of variables is automatically a relativistic theory, since a change of variables cannot alter this fact. From this perspective, one could consider even more general changes of variables than those of non-linear mappings between momentum variables of a single particle. In order to generate a non-trivial multiparticle sector, one could make a change of variables that mix the momenta of a number $N$ of momentum variables. Since in the case of a modified kinematics at first order, the composition of two momenta determines the composition of an arbitrary number of momenta, at this order it suffices to make this `generalized' change of variables involving the momenta of just two particles ($N=2$). In Sec.~\ref{sec:firstorder} of this paper, we will show that such a `generalized' change of momentum variables is able to reduce \textit{any} modified kinematics (with arbitrary composition law) to the SR kinematics at first order. In particular, this means that one could have obtained the results of Refs.~\cite{AmelinoCamelia:2011yi,Carmona2012} (the relations among the coefficients of the composition law, boosts and dispersion relation in a modified relativistic theory at first order) just by starting from standard SR and applying a generalized change of momentum variables, which is simpler than trying to impose the relativity principle itself in a generic deformed theory.  

This mathematical trick can be indeed of real help for the next step in the task of finding all possible DSR theories, which is considering second-order corrections in the modified kinematics, something much more involved (it contains many more coefficients) than in the first-order case. Generic kinematics at second order have been much less considered in the literature, not only for their mathematical complexity, but also because of their (generally assumed in the literature) less relevance in the phenomenological analysis of present and near-future experiments if first-order corrections are also present~\cite{AmelinoCamelia:2008qg}. This however could well be not the case: recent analyses of possible photon delays coming from gamma-ray bursts and blazars~\cite{Ackermann:2009aa,*HESS:2011aa,*Nemiroff:2011fk,*Vasileiou:2013vra,*Vasileiou:2015wja} seem to suggest the absence of first-order corrections as the leading quantum gravity effects (they constrain them, pushing the scale of $M$ to values larger than the Planck mass); also, the possible dropoff in the cosmological neutrino spectra observed by IceCube could be explained in terms of second-order, but not first-order, Planck-scale physics~\cite{Stecker2015}. These phenomenological indications, though still far from full confirmation, give the study of DSR theories at second order more relevance than that of an issue of simple academic interest.

In Sec.~\ref{sec:second} of the paper we will make use of a generalized change of momentum variables in order to study relativistic kinematics beyond SR at second order, and determine the relations between coefficients of the modified kinematics. Since the expressions are quite involved, we have found convenient to use a covariant notation, with a change of variables depending on the components of a fixed four-vector. As we will explicitly check in Sec.~\ref{sec:firstorder}, choosing afterwards this four-vector as a time-like vector, one recovers the results for rotational invariant deformations of SR.
We will also discuss the arbitrariness in the assignation of momentum variables to the particles participating in a process. For this purpose it will be relevant to see the difference between an ordinary composition law and a covariant one with respect to their properties related to a change of variables. 

Then, in Sec.~\ref{sec:Hopf} we will make explicit the correspondence between our notation and the algebraic language of Hopf algebras. This will allow us to see how $\kappa$-Poincar\'e is in fact an example of the generic deformations considered here. Finally, we will return to the discussion on the physical meaning of the momentum variables and some proposals to be developed in future work.

\section{Beyond SR at first order}
\label{sec:firstorder}

\subsection{A summary of previous results}
\label{sec:summary}

Let us first remind the results that were obtained in Ref.~\cite{Carmona2012}. A generic deformation of SR at order $1/M$, with the restriction of being polynomical in the components of the four-momentum $(p_0,p_i)$, and rotationally invariant, will have a modified dispersion relation (MDR) parametrized by two adimensional coefficients $\alpha_1, \alpha_2$:
\begin{equation}
C(p)=p_0^2-\vec{p}^2+\frac{\alpha_1}{M}p_0^3+\frac{\alpha_2}{M}p_0\vec{p}^2=m^2\,,
\label{eq:MDR}
\end{equation}
and a modified composition law (MCL) parametrized by five adimensional coefficients $\beta_1, \beta_2, \gamma_1, \gamma_2, \gamma_3$:
\begin{equation}
\left[p\oplus q\right]_0 = p_0 + q_0 + \frac{\beta_1}{M} \, p_0 q_0 + \frac{\beta_2}{M} \, \vec{p}\cdot\vec{q} \quad \quad \quad \quad \left[p \oplus q\right]_i = p_i + q_i + \frac{\gamma_1}{M} \, p_0 q_i + \frac{\gamma_2}{M} \, p_i q_0
+ \frac{\gamma_3}{M} \, \epsilon_{ijk} p_j q_k
\label{eq:MCL}
\end{equation}
where $\epsilon_{ijk}$ is the Levi-Civita symbol, a totally antisymmetric tensor, and it is implemented the condition
\begin{equation}
(p \oplus q)|_{q=0} = p \quad \quad  \quad \quad (p \oplus q)|_{p=0} = q\,.
\label{eq:cl0}
\end{equation}

In a theory with a relativity principle, the $\alpha_i$ coefficients of the MDR and the $(\beta_i,\gamma_i)$ coefficients of the MCL are in fact related by the so-called `golden rules'
\begin{equation}
\alpha_1=-\beta_1 \quad \quad \quad \alpha_2=\gamma_1+\gamma_2-\beta_2\,.
\label{eq:GR}
\end{equation}
This is so because both the parameters of the dispersion relation and the parameters of the modified composition law can be written as combinations of the coefficients that appear in the non-linear Lorentz transformations, in a way completely determined by the relativity principle. In Ref.~\cite{Carmona2012} it was obtained the result
\begin{alignat}{3}
\alpha_1 &=-2(\lambda_1+\lambda_2+2\lambda_3) \quad\quad &\alpha_2&=2(\lambda_1+2\lambda_2+3\lambda_3) \quad\quad &  \label{alphalambda}\\
\beta_1 &= 2 \,(\lambda_1 + \lambda_2 + 2\lambda_3) \quad\quad & 
\beta_2 &= -2 \lambda_3 - \eta_1^L - \eta_1^R \quad\quad & \label{betalambda}\\
\gamma_1 &= \lambda_1 + 2\lambda_2 + 2\lambda_3 - \eta_1^L \quad\quad & 
\gamma_2 &= \lambda_1 + 2\lambda_2 + 2\lambda_3 - \eta_1^R \quad\quad & \gamma_3 = \eta_2^L - \eta_2^R \,,
\label{gammalambda}
\end{alignat}
where the $\lambda_i$ are coefficients of the one-particle boost, and $\eta^L_i, \eta^R_i$ are coefficients appearing in a boost transformation of the two-particle system. Specifically, the notation is as follows. In the one-particle system, the momentum $p$ transforms under an infinitesimal boost by means of a deformed Lorentz transformation $p\to T(p)$:
\begin{eqnarray}
\left[T(p)\right]_0 &=& p_0 + (\vec{p} \cdot \vec{\xi}) + \frac{\lambda_1}{M} \, p_0 (\vec{p} \cdot \vec{\xi}) \nonumber \\
\left[T(p)\right]_i &=& p_i + p_0 \xi_i + \frac{\lambda_2}{M} \, p_0^2 \xi_i + \frac{\lambda_3}{M} \, {\vec p}^{\,2} \xi_i + \frac{(\lambda_1 + 2\lambda_2 + 2\lambda_3)}{M} \, p_i ({\vec p} \cdot {\vec \xi}) \,,
\label{T-one}
\end{eqnarray}
where $\vec{\xi}$ is the vector parameter of the boost, and the $\lambda_i$ are dimensionless coefficients. The form of Eq.~\eqref{T-one} is obtained after one imposes the condition that the modified boosts must reproduce the Lorentz algebra, ie, that the commmutator of two boosts corresponds to a rotation~\cite{Carmona2012}, and the invariance of the dispersion relation under this boost, $C(T(p))=C(p)$, leads to Eq.~\eqref{alphalambda}.

The transformation law of a two-particle system is non-trivial. Since the non-linear terms of the MCL Eq.~\eqref{eq:MCL} mix the components of the momenta which are being composed and, for a generic MCL, the order of the momenta is relevant, a boost transformation on $p$ and $q$ will also mix both momenta in an order-dependent way. When one of the momenta is zero, however, the boost transformation of the other momentum is determined by Eq.~\eqref{T-one}, that is, under a boost, $(p,q) \to (T^L_q(p),T^R_p(q))$ such that
\begin{equation}
T^L_q(p) \,=\, T(p) + {\bar T}^L_q(p) {\hskip 1cm} T^R_p(q) \,=\, T(q) + {\bar T}^R_p(q) \,,
\label{eq:boost2}
\end{equation}
and rotational invariance gives the generic expressions%\footnote{Note that there is a change of sign in the definition of $\eta_2^R$ with respect to the definition used in Ref.~\cite{Carmona2012}.}
\begin{eqnarray}
\left[{\bar T}^L_q(p)\right]_0 &=& \frac{\eta_1^L}{M} \, q_0 ({\vec p} \cdot {\vec \xi}) + \frac{\sigma_1^L}{M} \, p_0 ({\vec q} \cdot {\vec \xi}) + \frac{\eta_2^L}{M} \, ({\vec p} \wedge {\vec q}) \cdot {\vec \xi} \nonumber \\
\left[{\bar T}^L_q(p)\right]_i &=&  \frac{\eta_3^L}{M} \, q_i ({\vec p} \cdot {\vec \xi}) + \frac{\sigma_2^L}{M} \, p_i ({\vec q} \cdot {\vec \xi}) + \frac{\eta_4^L}{M} \, q_0 \epsilon_{ijk} p_j \xi_k + \frac{\sigma_3^L}{M} \, ({\vec p} \cdot {\vec q}) \xi_i + \frac{\sigma_4^L}{M} \, p_0 \epsilon_{ijk} q_j \xi_k + \frac{\sigma_5^L}{M} \, p_0 q_0 \xi_i \nonumber \\
\left[{\bar T}^R_p(q)\right]_0 &=& \frac{\eta_1^R}{M} \, p_0 ({\vec q} \cdot {\vec \xi}) + \frac{\sigma_1^R}{M} \, q_0 ({\vec p} \cdot {\vec \xi}) + \frac{\eta_2^R}{M} \, ({\vec q} \wedge {\vec p}) \cdot {\vec \xi} \nonumber \\
\left[{\bar T}^R_p(q)\right]_i &=&  \frac{\eta_3^R}{M} \, p_i ({\vec q} \cdot {\vec \xi}) + \frac{\sigma_2^R}{M} \, q_i ({\vec p} \cdot {\vec \xi}) + \frac{\eta_4^R}{M} \, p_0 \epsilon_{ijk} q_j \xi_k + \frac{\sigma_3^R}{M} \, ({\vec q} \cdot {\vec p}) \xi_i + \frac{\sigma_4^R}{M} \, q_0 \epsilon_{ijk} p_j \xi_k + \frac{\sigma_5^R}{M} \, q_0 p_0 \xi_i
\label{eq:gen2pboost}
\end{eqnarray}
After imposing that the transformations must satisfy the Lorentz algebra, one gets the following relations between the previous coefficients:
\begin{align}
\eta_2^L=\sigma_4^L=-\eta_4^L \quad& \quad \eta_1^L-\eta_3^L+\sigma_3^L+\sigma_5^L=\sigma_1^L-\sigma_2^L+\sigma_3^L+\sigma_5^L=0 \nonumber \\
\eta_2^R=\sigma_4^R=-\eta_4^R \quad &\quad \eta_1^R-\eta_3^R+\sigma_3^R+\sigma_5^R=\sigma_1^R-\sigma_2^R+\sigma_3^R+\sigma_5^R=0 \,.
\label{eq:eta-sigma}
\end{align}
It turns out that, with such a generic two-particle transformation, the function $C(T^L_q(p))$, as well as $C(T^R_p(q))$, is not equal to $C(p)$, and in fact, both of the functions $C(T^L_q(p))$ and $C(T^R_p(q))$ depend on both $p$ and $q$, unless the transformation satisfies the additional restriction
\begin{align}
\sigma_2^L=0 \quad \quad \sigma_3^L&=-\eta_3^L \quad \quad \sigma_5^L=\eta_1^L \nonumber \\
\sigma_2^R=0 \quad \quad \sigma_3^R&=-\eta_3^R \quad \quad \sigma_5^R=\eta_1^R  \,.
\label{eq:MDR-restriction}
\end{align}
This is a necessary restriction in an extension of the SR kinematics based on the same two ingredients that are used when analyzing the kinematics of a process in SR: a conservation law derived from a (modified) composition law, and a (modified) dispersion relation for each of the particles participating in the process, where, by definition, the dispersion relation of a particle is a function of the momentum of that particle. Algebraically, this means that we ask that the Casimir of the Lorentz algebra in the one-particle system must be also a Casimir in the two-particle system. In this sense we have a non-linear representation of the Lorentz algebra in a momentum space that is consistently extended to the tensor product of momentum spaces. As we will see in Sect.\ref{sec:Hopf}, this condition is automatically satisfied in the formalism of Hopf algebras.

Relations \eqref{eq:MDR-restriction} and~\eqref{eq:eta-sigma} imply that a generic boost in the two-particle system has only four free parameters, that can be chosen as $\eta_1^L$,  $\eta_2^L$, $\eta_1^R$, and $\eta_2^R$, since the rest of the coefficients in Eq.~\eqref{eq:gen2pboost} are expressible in terms of them as
\begin{equation}
\sigma_1^{L,R}=0, \quad \sigma_2^{L,R}=0, \quad \sigma_3^{L,R}=-\eta_1^{L,R}, \quad \sigma_4^{L,R}=\eta_2^{L,R}, \quad \sigma_5^{L,R}=\eta_1^{L,R}, \quad
\eta_3^{L,R}=\eta_1^{L,R}, \quad \eta_4^{L,R}=-\eta_2^{L,R}\,.
\label{eq:}
\end{equation}
Finally, the relativity principle imposed as the invariance of the composition of momenta for different inertial observers, that is,
\begin{equation}
T(p\oplus q)=T_q^L(p)\oplus T_p^R(q) \,,
\label{eq:RP-1}
\end{equation} 
leads, after some tedious algebra, to Eqs.~\eqref{betalambda} and\eqref{gammalambda}.

\subsection{Change of variables and change of basis}
\label{sec:change}

In this section we are going to show that the final result Eqs.~\eqref{alphalambda}-\eqref{gammalambda} can be obtained in a much easier way with the help of a mathematical trick: a change of variables in the two-particle system.

In Ref.~\cite{Carmona2012} we already showed that a generic (again, rotational invariant) change of energy-momentum variables $(p_0,\vec{p}) \to (P_0,\vec{P})$ of the form
\begin{equation}
\begin{split}
p_0& =P_0+\frac{\delta_1}{M}P_0^2+\frac{\delta_2}{M}\vec{P}^2\equiv \mathcal{B}_0(P_0,\vec{P})\equiv \mathcal{B}_0(P) \\
p_i&=P_i+\frac{\delta_3}{M} P_0 P_i \equiv \mathcal{B}_i(P),
\end{split}
\label{eq:ch-base}
\end{equation} 
is able to change the coefficients $\beta_1$, $\beta_2$ and $(\gamma_1+\gamma_2)$ of the MCL Eq.~\eqref{eq:MCL} (the only parameters in a commutative MCL at first order) as a function of the parameters $\delta_1$, $\delta_2$ and $\delta_3$. Following the mathematical language of Hopf algebras (see Sec.~\ref{sec:Hopf}), we will refer to Eq.~\eqref{eq:ch-base} as a \emph{change of basis} in momentum space. That is: a change of basis is a change of momentum variables that is the same for all particles in a process. In the geometric language of relative locality, a change of basis would be just a choice of coordinates in momentum space. As we explained in the Introduction, a change of basis has not got any algebraic or geometric content in both formalisms. From the point of view of DSR, however, the momentum variables which are changed under a change of basis are physically inequivalent, although it is difficult to give a physical meaning to these variables in the absence of a dynamical theory or of a theory of space-time (see more comments on this in Sec.~\ref{sec:concl}). For the moment, in Ref.~\cite{Carmona2012} we just stated that a change of basis can relate a generic commutative MCL at first order with the additive composition law of SR.

There is a technical point that was not made clear enough in Ref.~\cite{Carmona2012}. If $(P_0,\vec{P})$ are energy-momentum variables that compose additively, which is the modified composition law for $(p_0,\vec{p})$ that is derived from the change of basis~\eqref{eq:ch-base}? The simple definition $(p\oplus q)_\mu\equiv (P+Q)_\mu$ does not work, since, using the inverse of the change of basis
\begin{equation}
\begin{split}
P_0& =p_0-\frac{\delta_1}{M}p_0^2-\frac{\delta_2}{M}\vec{p}^2\equiv \mathcal{B}^{-1}_0(p) \\
P_i&=p_i-\frac{\delta_3}{M} p_0 p_i \equiv \mathcal{B}^{-1}_i(p),
\end{split}
\label{eq:invch-base}
\end{equation}
we see that
\begin{equation}
(P+Q)_0=P_0+Q_0=\mathcal{B}^{-1}_0(p)+\mathcal{B}^{-1}_0(q)=p_0+q_0-\frac{\delta_1}{M} (p_0^2+q_0^2)-\frac{\delta_2}{M}(\vec{p}^2+\vec{q}^2),
\label{eq:MCL-1}
\end{equation}
and this cannot be used to define $(p\oplus q)_0$, since the previous expression does not satisfy the condition given by Eq.~(\ref{eq:cl0}). Condition~(\ref{eq:cl0}) is however automatically guaranteed if we define the MCL as
\begin{equation}
(p\oplus q)_\mu \equiv \mathcal{B}_\mu\left(\mathcal{B}^{-1}(p)+\mathcal{B}^{-1}(q)\right)\,.
\label{eq:MCLdef}
\end{equation}
This is indeed what has been done in the literature when discussing the so-called `auxiliary variables' in the DSR framework~\cite{Judes:2002bw}. In this case, one gets
\begin{align}
(p\oplus q)_0& =p_0+q_0+\frac{2\delta_1}{M}p_0 q_0+\frac{2\delta_2}{M}\vec{p}\cdot\vec{q}\\
(p\oplus q)_i& =p_i+q_i+\frac{\delta_3}{M}p_0 q_i+\frac{\delta_3}{M}q_0 p_i\,.
\label{eq:MCL-2}
\end{align}
Comparing with Eq.~\eqref{eq:MCL}, we see that the composition law obtained by applying a change of basis to energy-momentum variables that compose additively is such that $\beta_1=2\delta_1$, $\beta_2=2\delta_2$, $\gamma_1=\gamma_2=\delta_3$, and $\gamma_3=0$. 

A change of basis does also change the transformation law of the energy-momentum variables. If $(P_0,\vec{P})$ transform linearly under Lorentz boosts, $P'_0=P_0+\vec{P}\cdot\vec{\xi}\,$, $\vec{P}'=\vec{P}+P_0\vec{\xi}\,$, then 
\begin{align}
[T(p)]_0 \equiv \mathcal{B}_0(P')&=P_0+\vec{P}\cdot\vec{\xi}+\frac{\delta_1}{M}(p_0+\vec{p}\cdot\vec{\xi})^2+\frac{\delta_2}{M}(\vec{p}+p_0\vec{\xi})^2=p_0+\frac{(2\delta_1+2\delta_2-\delta_3)}{M}p_0\vec{p}\cdot\vec{\xi} \\
[T(p)]_i \equiv \mathcal{B}_i(P')&=P_i+P_0\xi_i+\frac{\delta_3}{M}(p_0+\vec{p}\cdot\vec{\xi})(p_i+p_0\xi_i)=p_i+p_0 \xi_i+\frac{(\delta_3-\delta_1)}{M}p_0^2 \xi_i-\frac{\delta_2}{M}\vec{p}^2\xi_i+\frac{\delta_3}{M}p_i \vec{p}\cdot\vec{\xi} \, ,
\label{eq:ch-LT}
\end{align}
so that, comparing with Eq.~\eqref{T-one},
\begin{equation}
\lambda_1=2\delta_1+2\delta_2-\delta_3 \quad ,\quad \lambda_2=\delta_3-\delta_1 \quad , \quad \lambda_3=-\delta_2\,.
\label{eq:lambdafromdelta}
\end{equation}

The variables $(p_0,\vec{p})$, therefore, transform non-linearly. Accordingly, the dispersion relation in terms of these variables will be:
\begin{equation}
C(p)\equiv P_0^2-\vec{P}^2=p_0^2-\vec{p}^2-\frac{2\delta_1}{M}p_0^3+\frac{2(\delta_3-\delta_2)}{M}p_0\vec{p}^2=m^2,
\label{eq:ch-DR}
\end{equation}
and, comparing with Eq.~\eqref{eq:MDR},
\begin{equation}
\alpha_1=-2\delta_1=-\beta_1 \quad \quad \alpha_2=2(\delta_3-\delta_2)=\gamma_1+\gamma_2-\beta_2.
\label{eq:alphafromdelta}
\end{equation}
This agrees both with Eqs.~\eqref{eq:lambdafromdelta} and~\eqref{alphalambda}, and with the golden rules Eq.~\eqref{eq:GR}.

We will now define what we will call a \emph{change of variables} as opposed to simply a change of basis. By definition, a change of variables $(P,Q) \to (p,q)$ will mix momenta of the two particle system, in such a way that when one of the momenta is equalled to zero, then the change of variables is just the identity function: 
\begin{equation}
(P,Q) \to (p,q)=(\mathcal{F}^L(P,Q),\mathcal{F}^R(P,Q)) \,\text{ such that } \mathcal{F}^L(P,0)=P ,\, \mathcal{F}^L(0,Q)=0 \text{ and } \mathcal{F}^R(0,Q)=Q ,\, \mathcal{F}^R(P,0)=0\,.
\label{eq:restrictedch}
\end{equation}

Such a transformation has not a simple interpretation in the algebraic or geometric languages, unlike the case of a change of basis. We will use it however as a mathematical trick to generate a modified relativistic kinematics from variables that transform linearly in the two-particle momentum space; in fact, we will see that any relativistic kinematics at first order can be generated from a change of variables and a change of basis applied to the standard variables of SR, so that one can use this procedure to get the results of Sec.~\ref{sec:summary} in a much easier way.

Let us first take $(P,Q)$ variables that transform linearly and compose additively (that is, they are the standard variables of SR). Then, although the momentum variables $(p,q)$ transform non-linearly in the two-particle system ($T^L_q(p)$ and $T^R_p(q)$ are not linear), $T(p)$ is a linear transformation, since when $q=0$, $p=P$. In the notation used above, this means that the $\lambda_i$ are equal to zero. Since we can generate $\lambda_i\neq 0$ from a change of basis, we will say that the $\lambda_i$ are zero in the \emph{classical basis} of momentum space (this is a nomenclature inherited from the Hopf algebras framework, see Sec.~\ref{sec:Hopf}). A change of variables from SR variables, therefore, will generate a relativistic kinematics beyond SR for momentum variables in the classical basis.

It is easy to check that a change of variables generate a modified composition law and transformation laws that are compatible with the relativity principle. To see this, first note that the natural definition
\begin{equation}
p\oplus q \equiv P+Q
\label{eq:MCL-ch}
\end{equation}
is indeed a good definition of a modified composition law, since it satisfies the property Eq.~\eqref{eq:cl0}. For example,
\begin{equation}
(p \oplus q)|_{q=0} =\left[\left(\mathcal{F}^{-1}\right)^L (p,q)+ \left(\mathcal{F}^{-1}\right)^R (p,q)\right]_{q=0}=p+0=p\,,
\label{eq:demo1}
\end{equation}
where we have made use of the inverse of the change of variables, $P=\left(\mathcal{F}^{-1}\right)^L (p,q)$, $Q=\left(\mathcal{F}^{-1}\right)^R (p,q)$ and the properties indicated in Eq.~\eqref{eq:restrictedch}, which express that when $Q=0 \Rightarrow p=P, \, q=0$; that is, $\left(\mathcal{F}^{-1}\right)^L (p,0)=p$, $\left(\mathcal{F}^{-1}\right)^R (p,0)=0$.
Note that from Eq.~\eqref{eq:MCL-ch}, it is evident that $p\oplus q$ transform linearly, that is, as a single momentum does.

Secondly, the transformations properties of $(p,q)$ are defined from 
\begin{equation}
T_q^L(p)\equiv\mathcal{F}^L(P',Q')  \quad \quad T_p^R(q)\equiv\mathcal{F}^R(P',Q')\,,
\label{eq:MTL-ch}
\end{equation}
where $P'$, $Q'$, denote the usual linear Lorentz transformations. Then,
\begin{equation}
T_q^L(p) \oplus T_p^R(q)= \mathcal{F}^L(P',Q') \oplus \mathcal{F}^R(P',Q') = P'+Q' = (P+Q)' = T(p\oplus q)\, ,
\label{eq:demo2}
\end{equation}
where we have used Eq.~\eqref{eq:MTL-ch}, the fact that $P$ and $Q$ transform linearly, and the definition Eq.~\eqref{eq:MCL-ch}. Condition~\eqref{eq:RP-1} is, therefore, automatically guaranteed by a change of variables.

Finally, since $T(p)$ is the linear Lorentz transformation, the Casimir is just the usual one, $C(p)=p_0^2-\vec{p}^2=m^2$ (this is of course a property of the classical basis). As explained in Sec.~\ref{sec:summary}, we must make sure that this is also the Casimir of the two-particle system. In terms of the change of variables, we must impose that $P_0^2-\vec{P}^2=p_0^2-\vec{p}^2$, $Q_0^2-\vec{Q}^2=q_0^2-\vec{q}^2$. A generic (rotational invariant) change of variables that satisfies this condition has four parameters at first order in $1/M$:
\begin{equation}
\begin{split}
P_{0}=p_{0}+\frac{v_{1}^{L}}{M}\vec{p}.\vec{q}\qquad &  P_{i}=p_{i}+\frac{v_{1}^{L}}{M}p_{0}q_{i}+\frac{v_{2}^{L}}{M}\epsilon_{ijk}p_{j}q_{k}
\\
Q_{0}=q_{0}+\frac{v_{1}^{R}}{M}\vec{p}.\vec{q}\qquad & Q_{i}=q_{i}+\frac{v_{1}^{R}}{M}q_{0}p_{i}+\frac{v_{2}^{R}}{M}\epsilon_{ijk}q_{j}p_{k}.
\label{ch-var}
\end{split}
\end{equation}
The composition law in these variables is 
\begin{equation}
\begin{split}
\left[p\oplus q\right]_0&=P_0+Q_0=p_{0}+q_{0}+\frac{v_{1}^{L}+v_{1}^{R}}{M}\vec{p}.\vec{q}
\\
\left[p\oplus q\right]_i&=P_i+Q_i=p_{i}+q_{i}+\frac{v_{1}^{L}}{M}p_{0}q_{i}+\frac{v_{1}^{R}}{M}q_{0}p_{i}+\frac{v_{2}^{L}-v_{2}^{R}}{M}\epsilon_{ijk}p_{j}q_{k}
\end{split}
\label{chvar-cl-1st}
\end{equation}
so that comparison with Eq.~\eqref{eq:MCL} gives
\begin{align}
\beta_{1}\,=\,0{\hskip1cm}\beta_{2}\,=\,v_{1}^{L}+v_{1}^{R}
{\hskip1cm}\gamma_{1}\,=\,v_{1}^{L}{\hskip1cm}\gamma_{2}\,=\,v_{1}^{R}{\hskip1cm}\gamma_{3}\,=\,v_{2}^{L}-v_{2}^{R}.
\label{MCLpar-ch}
\end{align}
This is in fact the general solution of the golden rules Eq.~\eqref{eq:GR} when $\alpha_1=\alpha_2=0$ (classical basis).
We can then redefine the golden rules as a set of relations that the coefficients of the composition law must satisfy when working in the classical basis:
\be
\beta_1=0 \quad \quad \beta_2=\gamma_1+\gamma_2 \,.
\label{eq:newGR}
\ee
For the transformation law one obtains
\begin{align}
\eta_{1}^{L,R}=-v_{1}^{L,R}\qquad \eta_{2}^{L,R}=v_{2}^{L,R}\,.
\label{eta-v}
\end{align}
Eqs.~\eqref{betalambda} and~\eqref{gammalambda} are then immediately obtained for the case $\lambda_1=\lambda_2=\lambda_3=0$. One can get the full relations by combining a change of variables with a change of basis.

We observe that there is a one to one correspondence between the parameters of the change of variables~\eqref{ch-var} and the $\eta$ parameters of a non-linear boost in the two-particle system, on the one hand, and the parameters of the change of basis~\eqref{eq:ch-base} and the $\lambda$ parameters of a non-linear boost in the one-particle system, on the other hand. A change of variables and a change of basis from linear variables (that is, from variables that transform as in SR), therefore, produce the most generic relativistic kinematics beyond SR at first order. We will assume that this property is true at every order, so that generic relativistic kinematics at arbitrary order can be easily analyzed by this mathematical procedure.\footnote{We do not have a proof of this assumption. However, even if this were not the case, the richness of the relativistic theories that can be obtained by this procedure merits a study on its own. In fact, one should extend the change of variables presented in this section to a change of variables of a number $N$ of momenta when studying relativistic theories beyond SR at order $(N-1)$, but in order to reduce the complexity of the study, we will make the additional simplification of considering the change of variables just in the two-particle system, which has a consequence on the type of modified composition laws that can be generated, see Secs.~\ref{sec:specific} and~\ref{sec:second}.}
 
\subsection{Covariant notation}
\label{sec:covariant}

It is convenient to introduce a covariant notation that will simplify the calculations at second order. We consider again a change of variables in the two-particle system, $(P, Q) \to (p, q)$, from linear variables $(P,Q)$ to nonlinear variables $(p,q)$, which, as explained in the previous section, must satisfy
\be
(P, 0) \to (P, 0) {\hskip 1cm} (0, Q) \to (0, Q) {\hskip 1cm} P^2 = p^2 {\hskip 1cm} Q^2 = q^2 \,.
\ee
The condition that the Casimir in the one-particle system has to be also a Casimir in the two-particle system, or equivalently, that the dispersion relation does not mix the $p,q$ variables when applying the change of variables to the standard dispersion relation, $P_0^2-\vec{P}^2=m^2$, corresponds to the fact that the terms in the power expansion proportional to $(1/M)$ must be orthogonal to $p$ in the expression of $P$, or to $q$ in the expression of $Q$, in order to have $p^2=P^2$, $q^2=Q^2$ at first order. A general change of variables in covariant notation is then of the form
\begin{align}
P_\mu &= p_\mu + \frac{v_1^L}{M} \left[q_\mu (n\cdot p) - n_\mu (p\cdot q)\right] + \frac{v_2^L}{M} \epsilon_{\mu\nu\rho\sigma} p^\nu q^\rho n^\sigma \\
Q_\mu &= q_\mu + \frac{v_1^R}{M} \left[p_\mu (n\cdot q) - n_\mu (p\cdot q)\right] + \frac{v_2^R}{M} \epsilon_{\mu\nu\rho\sigma} q^\nu p^\rho n^\sigma \,,
\label{p,q}
\end{align}
where we have made the change of variables depend on a fixed dimensionless vector $n$, which breaks Lorentz invariance, and $\epsilon_{0123}=-1$. The rotational invariant change of variables~\eqref{ch-var} is then obtained when $n_\mu=(1,0,0,0)$.~\footnote{When one has a formally covariant expression (it would be really covariant if $n$ were a vector transforming linearly under Lorentz transformations instead of a fixed vector) with an arbitrary fixed vector $n$, then the change of variables is covariant under the subgroup of Lorentz transformations that leaves the direction of $n$ invariant.} 

The change of variables modifies the composition law of the $(P,Q)$ variables, which is additive, $[P\bigo Q]_\mu = P_\mu + Q_\mu$, to a nonadditive composition law for the $(p,q)$ variables, $(p\oplus q)$, such that
\be
\begin{split}
\left(p\oplus q\right)_\mu \equiv  \left[P \bigo Q\right]_\mu = P_\mu + Q_\mu & = p_\mu + q_\mu + \frac{v_1^L}{M} (n\cdot p) q_\mu + \frac{v_1^R}{M} (n\cdot q) p_\mu \\ & - \frac{(v_1^L+v_1^R)}{M} n_\mu (p\cdot q) +
\frac{(v_2^L-v_2^R)}{M} \epsilon_{\mu\nu\rho\sigma} p^\nu q^\rho n^\sigma \,.
\end{split}
\label{cl1}
\ee
If we take $n_\mu = (1, 0, 0, 0)$ in the previous expression, then we have
\be
\begin{split}
\left[p\oplus q\right]_0 &= p_0 + q_0 + \frac{(v_1^L + v_1^R)}{M} \vec{p}\cdot \vec{q} \\
\left[p\oplus q\right]_i &= p_i + q_i + \frac{v_1^L}{M} p_0 q_i + \frac{v_1^R}{M} q_0 p_i + \frac{(v_2^L - v_2^R)}{M} \epsilon_{ijk} p_j q_k \,,
\end{split}
\ee
which is the result given in Eq.~\eqref{chvar-cl-1st}.  

Let us consider now the Lorentz transformations. We will simplify the notation of Secs.~\ref{sec:summary} and~\ref{sec:change} and, instead of using $T_q^L(p)$or $T_p^R(q)$ we will denote by $(p',q')$ or $(P',Q')$ to the transformed momenta of $(p,q)$ or $(P,Q)$, independently of whether the transformation is linear or not (but remember that in the case of a nonlinear transformation, $p'$ and $q'$ depend on both $p$ and $q$). We also introduce the notation 
$\tilde{X}_\mu \equiv \Lambda_\mu^{\:\nu} X_\nu$, where the $\Lambda_\mu^{\:\nu}$ are the standard Lorentz transformation matrices. Then we have 
\be
\begin{split}
p'_\mu + \frac{v_1^L}{M} \left[q'_\mu (n\cdot p') - n_\mu (p'.q')\right] + \frac{v_2^L}{M} \epsilon_{\mu\nu\rho\sigma} p'^\nu q'^\rho n^\sigma &\equiv P'_\mu  \\
& = \Lambda_\mu^{\:\nu} P_\nu = \tilde{p}_\mu + \frac{v_1^L}{M} \left[\tilde{q}_\mu (\tilde{n}\cdot\tilde{p}) - \tilde{n}_\mu (\tilde{p}\cdot\tilde{q})\right] + \frac{v_2^L}{M} \epsilon_{\mu\nu\rho\sigma} \tilde{p}^\nu \tilde{q}^\rho \tilde{n}^\sigma \,.
\end{split}
\ee
This relation is telling us that $p'$ is equal to $\tilde{p}$ up to terms proportional to $(1/M)$~\footnote{This is just an obvious consequence of the fact that the nonlinear Lorentz transformations of the new variables is due to the change of variables.}; then we have to first order
\be
p'_\mu = \tilde{p}_\mu + \frac{v_1^L}{M} \left[\tilde{q}_\mu \left((\tilde{n}-n)\cdot\tilde{p}\right) - \left(\tilde{n}_\mu - n_\mu\right) (\tilde{p}\cdot\tilde{q})\right] + \frac{v_2^L}{M} \epsilon_{\mu\nu\rho\sigma} \tilde{p}^\nu \tilde{q}^\rho (\tilde{n}^\sigma - n^\sigma) \,.
\ee
If we consider an infinitesimal Lorentz transformation, then
\be
\tilde{X}_\mu = X_\mu + \omega^{\alpha\beta} \eta_{\mu\alpha} X_\beta \,,
\ee
where $\omega^{\alpha\beta}=-\omega^{\beta\alpha}$ are the parameters of the infinitesimal transformation, and one has
\be
p'_\mu = \tilde{p}_\mu + \omega^{\alpha\beta} n_\beta \left[\frac{v_1^L}{M} \left(q_\mu p_\alpha - \eta_{\mu\alpha} (p\cdot q)\right) + \frac{v_2^L}{M} \epsilon_{\mu\alpha\nu\rho} p^\nu q^\rho\right] \,.
\label{p'1}
\ee
For the second variable $q$, similar arguments lead to 
\be
q'_\mu \,=\, \tilde{q}_\mu + \omega^{\alpha\beta} n_\beta \left[\frac{v_1^R}{M} \left(p_\mu q_\alpha - \eta_{\mu\alpha} (p\cdot q)\right) + \frac{v_2^R}{M} \epsilon_{\mu\alpha\nu\rho} q^\nu p^\rho\right]
\label{q'1}
\ee
that, together with (\ref{p'1}), gives the nonlinear Lorentz transformation of the variables $(p, q)$ at first order.   
When $n_\mu = (1, 0, 0, 0)$ one has
\be
\omega^{0\beta} n_\beta = 0 {\hskip 1cm} \omega^{i\beta} n_\beta = \omega^{i0}=\xi^i = -\xi_i \,,
\ee
and the Lorentz transformation at first order (\ref{p'1}), (\ref{q'1}) becomes
\be
\begin{split}
p'_0 &= p_0 + (\vec{p} \cdot \vec{\xi}) - \frac{v_1^L}{M} q_0 (\vec{p} \cdot \vec{\xi}) + \frac{v_2^L}{M} ({\vec p} \wedge {\vec q}) \cdot {\vec \xi} \\ 
p'_i &= p_i + p_0 \xi_i - \frac{v_1^L}{M} \left(q_i (\vec{p} \cdot \vec{\xi}) - (p\cdot q) \xi_i\right) - \frac{v_2^L}{M} \left(q_0 (\vec{p} \wedge \vec {\xi})_i - p_0 (\vec{q} \wedge \vec {\xi})_i\right) \\
q'_0 &= q_0 + (\vec{q} \cdot \vec{\xi}) - \frac{v_1^R}{M} p_0 (\vec{q} \cdot \vec{\xi}) + \frac{v_2^R}{M} ({\vec q} \wedge {\vec p}) \cdot {\vec \xi} \\ 
q'_i &= q_i + q_0 \xi_i - \frac{v_1^R}{M} \left(p_i (\vec{q} \cdot \vec{\xi}) - (p\cdot q) \xi_i\right) - \frac{v_2^R}{M} \left(p_0 (\vec{q} \wedge \vec {\xi})_i - q_0 (\vec{p} \wedge \vec {\xi})_i\right) \,,
\end{split}
\label{MLT-1st}
\ee
which gives again Eq.~\eqref{eta-v}.

This completes the discussion of a change of variables from linear variables at first order in covariant notation. As remarked, under the change of variables, the dispersion relation of $p$ and $q$ are the standard ones. If one wants to consider a modified dispersion relation, then it is necessary to introduce a change of basis
\be
X_\mu = \hat{X}_\mu + \frac{b_1}{M} \hat{X}_\mu (n\cdot\hat{X}) + \frac{b_2}{M} n_\mu \hat{X}^2 + \frac{b_3}{M} n_\mu (n\cdot\hat{X})^2
\label{eq:basiscov}
\ee
(where $X$ stands for $p$ or $q$), so that one has a first order dispersion relation
\be
m^2 = p^2 = \hat{p}^2 + \frac{2(b_1+b_2)}{M} \hat{p}^2 (n\cdot\hat{p}) + \frac{2 b_3}{M} (n\cdot\hat{p})^3 \,,
\label{drhat1}
\ee
and a new composition law at first order $\hat{p}\ohat\hat{q}$ that follows from Eqs.~\eqref{eq:MCLdef} and~\eqref{eq:basiscov}, that is,
\be
\left[p\oplus q\right]_\mu = \left[\hat{p}\ohat\hat{q}\right]_\mu + \frac{b_1}{M} (\hat{p}+\hat{q})_\mu \left(n\cdot(\hat{p}+\hat{q})\right) + \frac{b_2}{M} n_\mu (\hat{p}+\hat{q})^2 + \frac{b_3}{M} n_\mu \left(n\cdot(\hat{p}+\hat{q})\right)^2 \,.
\ee 
The final result for the new composition law is
\be
\begin{split}
\left[\hat{p}\ohat \hat{q}\right]_\mu & = \hat{p}_\mu + \hat{q}_\mu + \frac{v_1^L - b_1}{M} (n\cdot\hat{p}) \hat{q}_\mu + \frac{v_1^R - b_1}{M} (n\cdot\hat{q}) \hat{p}_\mu \\ & - \frac{(v_1^L+v_1^R) + 2 b_2}{M} n_\mu (\hat{p}\cdot\hat{q}) - \frac{2 b_3}{M} n_\mu (n\cdot\hat{p}) (n\cdot\hat{q}) + \frac{(v_2^L-v_2^R)}{M} \epsilon_{\mu\nu\rho\sigma} \hat{p}^\nu \hat{q}^\rho n^\sigma \,.
\end{split}
\label{clhat1}
\ee   
If we choose $n_\mu=(1, 0, 0, 0)$ in Eqs.~\eqref{drhat1} and~\eqref{clhat1}, we get 
\be
m^2 = \hat{p}^2 + \frac{2(b_1+b_2)}{M} \hat{p}^2 \hat{p}_0 + \frac{2 b_3}{M} \left(\hat{p}_0\right)^3 = 
\hat{p}_0^2 - {\vec{\hat{p}}}^2 + \frac{2(b_1+b_2+b_3)}{M} \left(\hat{p}_0\right)^3 - \frac{2(b_1+b_2)}{M}\hat{p}_0 \vec{\hat{p}}^2 \,,
\ee
to be compared with the general expression Eq.~\eqref{eq:MDR}, and
\be
\begin{split}
& \left[\hat{p}\ohat\hat{q}\right]_0 = \hat{p}_0 + \hat{q}_0 - \frac{2(b_1+b_2+b_3)}{M} \hat{p}_0 \hat{q}_0 + \frac{(v_1^L + v_1^R) + 2 b_2}{M}\, \vec{\hat{p}}\cdot \vec{\hat{q}} \\
& \left[\hat{p}\ohat \hat{q}\right]_i = \hat{p}_i + \hat{q}_i
+ \frac{v_1^L - b_1}{M} \hat{p}_0 \hat{q}_i + \frac{v_1^R - b_1}{M} \hat{q}_0 \hat{p}_i + \frac{(v_2^L - v_2^R)}{M} \epsilon_{ijk} \hat{p}_j \hat{q}_k \,,
\end{split}
\ee
to be compared with Eq.~\eqref{eq:MCL}, so that, indeed, the golden rules Eq.~\eqref{eq:GR} are satified, and we reproduce the results described in Sec.~\ref{sec:summary} for a general first order modification of the dispersion relation and composition law compatible with the relativity principle, with a linear implementation of rotational symmetry.

\subsection{Specific features at first order}
\label{sec:specific}

Before embarking on the analysis of second-order kinematics beyond SR, let us remark some points that are specific to first-order kinematics and that do not necessarily extend trivially to second order.
\begin{enumerate}
\item At first order, the modified composition law of an arbitrary number of particles is obtained from the MCL of two particles, whose generic form is given by Eq.~\eqref{eq:MCL}. This is a consequence of the natural requirement that a MCL of $N$ particles has to reduce to the MCL of $N-1$ particles if one of the momenta is made zero, and the fact that at first order all the non-linear terms are quadratic in momenta. For example, at second order it will appear a new term in the composition of three momenta proportional to all three momenta which does not appear in the composition law of two momenta. One can nevertheless restrict the study of a relativistic kinematics to those cases in which the MCL for $N$ momenta is determined by the MCL of two momenta at order higher than one. This is a simplification that allows to confine the change of variables to the two-particle system, exactly as at first order. Since, besides of being simpler, it is an important case from the point of view of the algebraic formalism (where the main structure is the coproduct of the Hopf algebra, equivalent to the composition law of two momenta) and the geometric formalism (the composition law of two momenta defines the curvature of momentum space), we will stick to it in the analysis of the kinematics at second order. 
\item At first order, the MCL is associative. This can be easily proved in the covariant notation introduced in the previous section. Writing generically
\be
(p\oplus q)_\mu=p_\mu+q_\mu+\frac{1}{M}B_\mu^{\:\nu\rho}p_\nu q_\rho\, ,
\ee
we have
\begin{equation}
\begin{split}
[(p\oplus q)\oplus k]_\mu&=(p\oplus q)_\mu+k_\mu+\frac{1}{M}B_\mu^{\:\nu\rho}(p+q)_\nu k_\rho=p_\mu+q_\mu+k_\mu+\frac{1}{M}B_\mu^{\:\nu\rho}(p_\nu q_\rho+p_\nu k_\rho+q_\nu k_\rho) \\
&\equiv (p\oplus q\oplus k)_\mu=[p\oplus (q\oplus k)]_\mu
\label{eq:assoc}
\end{split}
\end{equation}
where the definition of $(p\oplus q\oplus k)$ is a consequence of the requirement mentioned in the previous item, that is, $(p\oplus q\oplus k)|_{k=0}=(p\oplus q)$ (and the equivalent conditions when one makes zero $p$ or $q$). At higher orders, however, a generic MCL is not associative.
\item Given a momentum $k$, a MCL defines its (left-handed, $\hat{k}_L$, or right-handed, $\hat{k}_R$) \emph{antipode}, such that $\hat{k}_L\oplus k=k\oplus \hat{k}_R=0$. At first order in $1/M$, $\hat{k}_L=\hat{k}_R\equiv \hat{k}$. Working again in covariant notation:
\begin{equation}
(\hat{k} \oplus k)_{\mu}=0 \Rightarrow \hat{k}_\mu=-k_\mu+\frac{1}{M}B_\mu^{\:\nu \rho}k_\nu k_\rho \Rightarrow (k\oplus \hat{k})_\mu=0\,.
\label{eq:antipode}
\end{equation}
At higher order, however,  $\hat{k}_L\neq \hat{k}_R$ in general.
\item At first order, the MCL defines the \emph{conservation law} of a process. For example, in the disintegration of a particle of momentum $k$ into two other particles, of momenta $p$ and $q$, $A(k)\to B(p)+C(q)$, two possible conservation laws are $k=p\oplus q$ and $\hat{k}\oplus p \oplus q=0$. If the variables of the modified kinematics are obtained from a change of variables of SR momenta, both conservation laws are relativistically invariant: since the composition $p\oplus q=P+Q$ transforms linearly, the first conservation law is the equality between two four-vectors that transform linearly, and since the composition $\hat{k}\oplus p \oplus q=0=-K+P+Q$ also transforms linearly, the second conservation law is the equality between a quantity that transforms linearly and zero, which is a relativistically invariant assertion. Both of the conservation laws generalize the SR conservation law, that can be written as $K=P+Q$ or $-K+P+Q=0$, and it is immediate to see that both conservation laws are equivalent at first order in $1/M$:
\begin{equation}
\hat{k}\oplus p \oplus q = \hat{k}\oplus(p\oplus q)=0 \Leftrightarrow k=p\oplus q\,, 
\label{eq:sameconslaw}
\end{equation} 
where we have used the associativity of the composition law and the definition of antipode. The two conservation laws, however, are not equivalent beyond first order in $1/M$. In the present paper we will be discussing the modified Lorentz transformations and modified composition laws; we will not discuss which is the most appropriate form of the conservation law, which will certainly be associated with the dynamical theory (probably a modified form of relativistic field theory) behind the considered extension of SR. We just note that the second form of the conservation law has some inherent ambiguities which are absent in the first conservation law: the definition of the antipode (which can be ``left'' or ``right'') and the choice of which antipodes are present in the conservation law, those of the momenta of the initial state, or those of the momenta of the final state.

\item Eqs.~\eqref{betalambda} and~\eqref{gammalambda} show that, at first order in $1/M$, the coefficients of the composition law are completely determined by the coefficients appearing in the non-linear Lorentz transformations. However, this is no longer the case at second order. As we have seen, the relations among the coefficients of the modified dispersion relation, modified composition law and modified Lorentz transformations, can be understood as a result of a change of variables and a change of basis from momentum variables transforming linearly. In the case of a modified kinematics at first order, there are four parameters in a general change of variables and three parameters in a general change of basis (see Eqs.~\eqref{ch-var} and~\eqref{eq:ch-base}, respectively). The three coefficients of the modified Lorentz transformation in the one-particle system, the $\lambda_i$, are exclusively determined by the coefficients of the change of basis, Eq.~\eqref{eq:lambdafromdelta}, and the four coefficients of the modified Lorentz transformation in the two-particle system, the $\eta_i$, are completely determined by the coefficients of the change of variables, Eq.~\eqref{eta-v}. Since the change of variables and the change of basis modify the composition law, it seems natural that the coefficients of the composition law be determined by the coefficients of the non-linear Lorentz transformations. However, there is a subtlety here. As we will see in the next Section, at order higher than one there exist non-linear composition laws that are compatible with linear Lorentz transformations. This will represent a new ingredient which is absent at first order; we will examine it carefully in Sec.~\ref{sec:choice}. 
\end{enumerate}

\section{Beyond SR at second order}
\label{sec:second}

According to what we learnt in the last section, we will generate modified kinematics at second order through a change of variables from momenta transforming linearly. A general kinematics can then be completed by considering a change of basis; we will however make emphasis on the change of variables over the change of basis because we want to compare the results obtained here with those appeared in the literature of Hopf algebras, where the basis is a matter of choice. In other words, and using the algebraic language, we will be working in the classical basis, where the one-particle momenta transform linearly.

\subsection{Change of variables up to second order}

As in Sec.~\ref{sec:covariant}, we need the most general expression of a change of variables at second order $(P, Q) \to (p, q)$ that is consistent with $p^2=P^2$, $q^2=Q^2$. To this purpose, we start by considering the terms proportional to $(1/M)^2$ that are obtained when one applies the first order change of variables Eq.~\eqref{p,q} to $p^2$, $q^2$:
\be
\begin{split}
P^2 & = p^2 + \frac{v_1^L v_1^L}{M^2} \left[q^2 (n\cdot p)^2 - 2 (p\cdot q)(n\cdot p)(n\cdot q) + (p\cdot q)^2n^2\right] \\ & + \frac{v_2^L v_2^L}{M^2} \left[p^2 q^2 n^2 + 2 (p\cdot q)(n\cdot p) (n\cdot q) - p^2 (n\cdot q)^2 - q^2 (n\cdot p)^2 - (p\cdot q)^2n^2\right] \\
Q^2 &= q^2 + \frac{v_1^R v_1^R}{M^2} \left[p^2 (n\cdot q)^2 - 2 (p\cdot q)(n\cdot p)(n\cdot q) + (p\cdot q)^2n^2\right] \\ & + \frac{v_2^R v_2^R}{M^2} \left[p^2 q^2 n^2 + 2 (p\cdot q) (n\cdot p) (n\cdot q) - p^2 (n\cdot q)^2 - q^2 (n\cdot p)^2 - (p\cdot q)^2n^2\right]\,.
\end{split}
\ee 
In order to have $p^2=P^2$ to second order then we need to consider a change of variables of the form
\be
\begin{split}
P_\mu &=  p_\mu + \frac{v_1^L}{M} \left[q_\mu (n\cdot p) - n_\mu (p\cdot q)\right] + \frac{v_2^L}{M} \epsilon_{\mu\nu\rho\sigma} p^\nu q^\rho n^\sigma - \frac{v_1^L v_1^L}{2 M^2} \left[n_\mu q^2 (n\cdot p) - 2 n_\mu (p\cdot q)(n\cdot q) + q_\mu (p\cdot q) n^2\right] \\
& - \frac{v_2^L v_2^L}{2 M^2} \left[p_\mu q^2 n^2 + 2 n_\mu (p\cdot q) (n\cdot q) - p_\mu (n\cdot q)^2 - n_\mu q^2 (n\cdot p) - q_\mu (p\cdot q) n^2\right] \\ 
& + \frac{v_3^L}{M^2} \left[p_\mu (n\cdot p) - n_\mu p^2\right] (n\cdot q) + \frac{v_4^L}{M^2} \left[q_\mu (n\cdot p) - n_\mu (p\cdot q)\right] (n\cdot p) + 
\frac{v_5^L}{M^2} \left[q_\mu (n\cdot p) - n_\mu (p\cdot q)\right] (n\cdot q) \\ & +
\frac{v_6^L}{M^2} (n\cdot p) \epsilon_{\mu\nu\rho\sigma} p^\nu q^\rho n^\sigma + 
\frac{v_7^L}{M^2} (n\cdot q) \epsilon_{\mu\nu\rho\sigma} p^\nu q^\rho n^\sigma \,.
\end{split}
\label{P->p}
\ee
For the second variable $Q$ we have
\be
\begin{split}
Q_\mu &=  q_\mu + \frac{v_1^R}{M} \left[p_\mu (n\cdot q) - n_\mu (p\cdot q)\right] + \frac{v_2^R}{M} \epsilon_{\mu\nu\rho\sigma} q^\nu p^\rho n^\sigma - \frac{v_1^R v_1^R}{2 M^2} \left[n_\mu p^2 (n\cdot q) - 2 n_\mu (p\cdot q)(n\cdot p) + p_\mu (p\cdot q) n^2\right] \\
& - \frac{v_2^R v_2^R}{2 M^2} \left[q_\mu p^2 n^2 + 2 n_\mu (p\cdot q) (n\cdot p) - q_\mu (n\cdot p)^2 - n_\mu p^2 (n\cdot q) - p_\mu (p\cdot q) n^2\right] \\ 
& + \frac{v_3^R}{M^2} \left[q_\mu (n\cdot q) - n_\mu q^2\right] (n\cdot p) + \frac{v_4^R}{M^2} \left[p_\mu (n\cdot q) - n_\mu (p\cdot q)\right] (n\cdot q) + 
\frac{v_5^R}{M^2} \left[p_\mu (n\cdot q) - n_\mu (p\cdot q)\right] (n\cdot p) \\ & +
\frac{v_6^R}{M^2} (n\cdot q) \epsilon_{\mu\nu\rho\sigma} q^\nu p^\rho n^\sigma + 
\frac{v_7^R}{M^2} (n\cdot p) \epsilon_{\mu\nu\rho\sigma} q^\nu p^\rho n^\sigma \,.
\end{split}
\label{Q->q}
\ee

There are, therefore, a total of 14 parameters $(v_1^L,\ldots,v_7^L;v_1^R,\ldots,v_7^R)$ for a generic change of variables at second order. To find the composition law for the variables $(p, q)$ we have to apply the previous change of variables to the composition law of the momentum variables $(P, Q)$ which transform linearly. But at second order there is a nonlinear composition law compatible with a linear Lorentz transformation (recall the comment in the last point of Sec.~\ref{sec:specific}):
\be
\left[P\bigo Q\right]_\mu = P_\mu + Q_\mu + \frac{c_1}{M^2} P_\mu Q^2 + \frac{c_2}{M^2} Q_\mu P^2 + \frac{c_3}{M^2} P_\mu (P\cdot Q) + \frac{c_4}{M^2} Q_\mu (P\cdot Q) \,.
\label{ccl2}
\ee
It is easy to see that this composition law is invariant under the standard Lorentz transformation, $\tilde{X}_\mu = \Lambda_\mu^{\:\nu} X_\nu$, since the $(1/M^2)$ terms are covariant; we will call it a \emph{covariant} composition law (the existence of covariant composition laws at second order was also noted in Ref.~\cite{Ivetic:2016qtz}). A generic non-linear composition law and Lorentz transformation in the two-particle system will then be obtained by applying the change of variables~\eqref{P->p}-\eqref{Q->q} to the generic covariant composition law~\eqref{ccl2}. For the new composition law, one obtains
\be
\begin{split}
\left[p\oplus q\right]_\mu &= p_\mu + q_\mu + \frac{v_1^L}{M} \left[q_\mu (n\cdot p) - n_\mu (p\cdot q)\right] + \frac{v_1^R}{M} \left[p_\mu (n\cdot q) - n_\mu (p\cdot q)\right] + \frac{(v_2^L-v_2^R)}{M} \epsilon_{\mu\nu\rho\sigma} p^\nu q^\rho n^\sigma \\ 
& + \frac{c_1}{M^2} p_\mu q^2 + \frac{c_2}{M^2} q_\mu p^2 + \frac{c_3}{M^2} p_\mu (p\cdot q) + \frac{c_4}{M^2} q_\mu (p\cdot q) \\ & - \frac{v_1^L v_1^L}{2 M^2} \left[n_\mu q^2 (n\cdot p) - 2 n_\mu (p\cdot q)(n\cdot q) + q_\mu (p\cdot q) n^2\right]   - \frac{v_1^R v_1^R}{2 M^2} \left[n_\mu p^2 (n\cdot q) - 2 n_\mu (p\cdot q)(n\cdot p) + p_\mu (p\cdot q) n^2\right] \\ & - \frac{v_2^L v_2^L}{2 M^2} \left[p_\mu q^2 n^2 + 2 n_\mu (p\cdot q) (n\cdot q) - p_\mu (n\cdot q)^2 - n_\mu q^2 (n\cdot p) - q_\mu (p\cdot q) n^2\right] \\ &  - \frac{v_2^R v_2^R}{2 M^2} \left[q_\mu p^2 n^2 + 2 n_\mu (p\cdot q) (n\cdot p) - q_\mu (n\cdot p)^2 - n_\mu p^2 (n\cdot q) - p_\mu (p\cdot q) n^2\right] \\ & + \frac{v_3^L}{M^2} \left[p_\mu (n\cdot p) - n_\mu p^2\right] (n\cdot q) + \frac{v_3^R}{M^2} \left[q_\mu (n\cdot q) - n_\mu q^2\right] (n\cdot p) \\ &  + \frac{v_4^L}{M^2} \left[q_\mu (n\cdot p) - n_\mu (p\cdot q)\right] (n\cdot p) + \frac{v_4^R}{M^2} \left[p_\mu (n\cdot q) - n_\mu (p\cdot q)\right] (n\cdot q) \\ & + \frac{v_5^L}{M^2} \left[q_\mu (n\cdot p) - n_\mu (p\cdot q)\right] (n\cdot q) + \frac{v_5^R}{M^2} \left[p_\mu (n\cdot q) - n_\mu (p\cdot q)\right] (n\cdot p) \\ & + \frac{(v_6^L - v_7^R)}{M^2} (n\cdot p) \epsilon_{\mu\nu\rho\sigma} p^\nu q^\rho n^\sigma + \frac{(v_7^L - v_6^R)}{M^2} (n\cdot q) \epsilon_{\mu\nu\rho\sigma} p^\nu q^\rho n^\sigma \,.
\end{split}
\label{cl2}
\ee
We see that a generic composition law generated by a change of variables at second order has coefficients that depend on 16 parameters: the four parameters of the original covariant composition law, Eq.~\eqref{ccl2}, and 12 combinations of the 14 parameters of the change of variables \eqref{P->p}-\eqref{Q->q}. In the interesting particular case of rotational invariance, we take $n_\mu=(1, 0, 0, 0)$ in Eq.~\eqref{cl2} and we find
\be
\begin{split}
\left[p\oplus q\right]_0 &= p_0 + q_0 + \frac{(v_1^L + v_1^R)}{M} \vec{p}\cdot\vec{q} + \frac{(2 c_1 - v_1^L v_1^L- 2v_3^R)}{2 M^2} p_0 q^2 + \frac{(2 c_2 - v_1^R v_1^R - 2 v_3^L)}{2 M^2} q_0 p^2 \\ &  + 
\frac{(2 c_3 + v_1^R v_1^R - v_2^R v_2^R - 2 v_4^L - 2 v_5^R)}{2 M^2} p_0 (p\cdot q) + \frac{(2 c_4 + v_1^L v_1^L - v_2^L v_2^L - 2 v_5^L - 2 v_4^R)}{2 M^2} q_0 (p\cdot q) \\ & + \frac{(v_2^R v_2^R + 2 v_3^L+ 2 v_4^L + 2 v_5^R)}{2 M^2} p_0^2 q_0 + \frac{(v_2^L v_2^L + 2 v_3^R + 2 v_5^L + 2 v_4^R)}{2 M^2} p_0 q_0^2 \\
\left[p\oplus q\right]_i &= p_i + q_i + \frac{v_1^L}{M} p_0 q_i + \frac{v_1^R}{M} q_0 p_i + \frac{(v_2^L - v_2^R)}{M} \epsilon_{ijk}p_{j} q_{k} + \frac{(2c_1 - v_2^L v_2^L)}{2 M^2} p_i q^2 + \frac{(2 c_2 - v_2^R v_2^R)}{2 M^2} q_i p^2 \\ &  + \frac{(2c_3 - v_1^R v_1^R + v_2^R v_2^R)}{2 M^2} p_i (p\cdot q) + \frac{(2c_4 - v_1^L v_1^L + v_2^L v_2^L)}{2 M^2} q_i (p\cdot q)  + \frac{(v_2^R v_2^R + 2 v_4^L)}{2 M^2} p_0^2 q_i + \frac{(v_2^L v_2^L + 2 v_4^R)}{2 M^2} p_i q_0^2 \\ &  + \frac{(v_3^L+ v_5^R)}{M^2} p_i p_0 q_0 + \frac{(v_3^R + v_5^L)}{M^2} q_i p_0 q_0 + \frac{(v_6^L-v_7^R)}{M^2} p_0\, \epsilon_{ijk}p_{j} q_{k} + \frac{(v_7^L-v_6^R)}{M^2} q_0\, \epsilon_{ijk}p_{j} q_{k} \,.
\end{split}
\label{generalCL}
\ee
where we can identify as in the first order case the following adimensional coefficients:
\begin{align}
\beta_1 & = 0 &  \beta_2 & = v_1^L + v_1^R & 2\beta_3 &= 2 c_1 - v_1^L v_1^L- 2v_3^R \nonumber \\
2\beta_4 &= 2 c_2 - v_1^R v_1^R - 2 v_3^L  & 2\beta_5 &=  2 c_3 + v_1^R v_1^R - v_2^R v_2^R - 2 v_4^L - 2 v_5^R & 2\beta_6 &= 2 c_4 + v_1^L v_1^L - v_2^L v_2^L - 2 v_5^L - 2 v_4^R  \nonumber \\
2\beta_7 &= v_2^R v_2^R + 2 v_3^L+ 2 v_4^L + 2 v_5^R & 2\beta_8 & =v_2^L v_2^L + 2 v_3^R + 2 v_5^L + 2 v_4^R & \gamma_1 & = v_1^L \nonumber \\
\gamma_2 & =v_1^R & \gamma_3 &=v_2^L - v_2^R & 2\gamma_4 &= 2c_1 - v_2^L v_2^L \nonumber \\
2\gamma_5 &= 2 c_2 - v_2^R v_2^R & 2\gamma_6 &=2c_3 - v_1^R v_1^R + v_2^R v_2^R & 2\gamma_7 &=2c_4 - v_1^L v_1^L + v_2^L v_2^L\nonumber \\
2\gamma_8 &=v_2^R v_2^R + 2 v_4^L & 2\gamma_9 &=v_2^L v_2^L + 2 v_4^R & \gamma_{10} & =v_3^L+ v_5^R \nonumber \\
\gamma_{11} &=v_3^R + v_5^L & \gamma_{12} &=v_6^L-v_7^R & \gamma_{13} &=v_7^L-v_6^R\,.
\label{MCLpar-ch-2nd}
\end{align}
These relations constitute the generalization to second order of Eq.~\eqref{MCLpar-ch}. As in the first-order case, in the classical basis (that is, when the dispersion relation and Lorentz transformations in the one-particle system are those of SR) the golden rules are a set of relations among the coefficients of the composition law. From Eq.~\eqref{MCLpar-ch-2nd}, we find that the the first-order rules given in Eq.~\eqref{eq:newGR} are generalized to second order to:
\be
\begin{split}
\beta_1=\beta_2 - \gamma_1 - \gamma_2 &=0\\
\beta_3 + \beta_6 - \gamma_4 -\gamma_7 + \gamma_9 +\gamma_{11} -\frac{\gamma_1^2}{2}&=0\\
\beta_4 + \beta_5 - \gamma_5 -\gamma_6 + \gamma_8 +\gamma_{10} -\frac{\gamma_2^2}{2}&=0\\
\beta_7 - \gamma_8 -\gamma_{10}=\beta_8 - \gamma_9 -\gamma_{11} &=0\,.
\end{split}
\label{gr-upto-2nd}
\ee

In the case of a modified Lorentz transformation in the two-particle system, $(p,q) \to (p',q')$, one can follow the steps that led to Eqs.~\eqref{p'1} and~\eqref{q'1} in Sec.~\ref{sec:covariant}; in this case, one gets the following expression for the transformed momentum $p'$:
\be
\begin{split}
p'_\mu &= \tilde{p}_\mu + \omega^{\alpha\beta} n_\beta \left[\frac{v_1^L}{M} p_\alpha q_\mu - \frac{v_1^L}{M} \eta_{\alpha\mu} (p\cdot q) - \frac{v_2^L}{M} \epsilon_{\alpha\mu\nu\rho} p^\nu q^\rho\right] \\ & + \omega^{\alpha\beta} n_\beta \left[-\frac{v_1^L v_1^R}{M^2} \left(q_\alpha p_\mu - \eta_{\alpha\mu} (p\cdot q)\right) (n\cdot p) - \frac{v_1^L v_2^R}{M^2} \epsilon_{\alpha\mu\nu\rho} p^\nu q^\rho (n\cdot p) - \frac{v_1^L v_1^L}{M^2} p_\alpha q_\mu (n\cdot q) \right. \\ &  \left. + \frac{v_1^L v_2^L}{M^2} \epsilon_{\alpha\nu\rho\sigma} q_\mu p^\nu q^\rho n^\sigma + \frac{v_1^L v_1^L}{M^2} \left(p_\alpha q^2 - q_\alpha (p\cdot q)\right) n_\mu + \frac{v_1^L v_1^R}{M^2} \left(q_\alpha p^2 - p_\alpha (p\cdot q)\right) n_\mu \right. \\ &  \left. - \frac{v_2^L v_1^L}{M^2} \epsilon_ {\alpha\mu\nu\rho} q^\nu n^\rho (p\cdot q) + \frac{v_2^L v_1^R}{M^2} \epsilon_ {\alpha\mu\nu\rho} p^\nu n^\rho (p\cdot q) + \frac{v_2^L v_2^L}{M^2} \left[q_\alpha \left(p_\mu (n\cdot q) - q_\mu (n\cdot p)\right) - \eta_{\alpha\mu} \left((n\cdot q)(p\cdot q) - (n\cdot p) q^2\right)\right] \right. \\ &  \left. - \frac{v_2^L v_2^R}{M^2} \left[p_\alpha \left(q_\mu (n\cdot p) - p_\mu (n\cdot q)\right) - \eta_{\alpha\mu} \left((n\cdot p)(p\cdot q) - (n\cdot q) p^2\right)\right] + \frac{v_2^L v_2^L}{M^2} q_\alpha p_\mu (n\cdot q) \right. \\ &  \left. - \frac{(v_1^L v_1^L - v_2^L v_2^L)}{2 M^2} \left(p_\alpha n_\mu + \eta_{\alpha\mu} (n\cdot p)\right) q^2 + \frac{(v_1^L v_1^L - v_2^L v_2^L - v_5^L)}{M^2} \left(q_\alpha n_\mu + \eta_{\alpha\mu} (n\cdot q)\right) (p\cdot q) \right. \\ &  \left. + \frac{v_3^L}{M^2} \left(q_\alpha (n\cdot p) + p_\alpha (n\cdot q)\right) p_\mu - \frac{v_3^L}{M^2} \left(q_\alpha n_\mu + \eta_{\alpha\mu} (n\cdot q)\right) p^2 + \frac{v_4^L}{M^2} 2 p_\alpha q_\mu (n\cdot p) - \frac{v_4^L}{M^2} \left(p_\alpha n_\mu + \eta_{\alpha\mu} (n\cdot p)\right) (p\cdot q) \right. \\ &  \left. + \frac{v_5^L}{M^2} \left(q_\alpha (n\cdot p) + p_\alpha (n\cdot q)\right) q_\mu + \frac{v_6^L}{M^2} \left[p_\alpha \epsilon_{\mu\nu\rho\sigma} p^\nu q^\rho n^\sigma - \epsilon_{\alpha\mu\nu\rho} p^\nu q^\rho (n\cdot p)\right] + \frac{v_7^L}{M^2} \left[q_\alpha \epsilon_{\mu\nu\rho\sigma} p^\nu q^\rho n^\sigma - \epsilon_{\alpha\mu\nu\rho} p^\nu q^\rho (n\cdot q)\right]\right] \,.
\end{split}
\label{p'2}
\ee
The Lorentz transformation for the second momentum variable ($q'$) is obtained from the expression for $p'$ by the interchanges $p\leftrightarrow q$, $v^i_L \leftrightarrow v^i_R$,
\be
\begin{split}
q'_\mu &= \tilde{q}_\mu + \omega^{\alpha\beta} n_\beta \left[\frac{v_1^R}{M} q_\alpha p_\mu - \frac{v_1^R}{M} \eta_{\alpha\mu} (p\cdot q) - \frac{v_2^R}{M} \epsilon_{\alpha\mu\nu\rho} q^\nu p^\rho\right] \\ & + \omega^{\alpha\beta} n_\beta \left[-\frac{v_1^R v_1^L}{M^2} \left(p_\alpha q_\mu - \eta_{\alpha\mu} (p\cdot q)\right) (n\cdot q) - \frac{v_1^R v_2^L}{M^2} \epsilon_{\alpha\mu\nu\rho} q^\nu p^\rho (n\cdot q) - \frac{v_1^R v_1^R}{M^2} q_\alpha p_\mu (n\cdot p) \right. \\ &  \left. + \frac{v_1^R v_2^R}{M^2} \epsilon_{\alpha\nu\rho\sigma} p_\mu q^\nu p^\rho n^\sigma + \frac{v_1^R v_1^R}{M^2} \left(q_\alpha p^2 - p_\alpha (p\cdot q)\right) n_\mu + \frac{v_1^R v_1^L}{M^2} \left(p_\alpha q^2 - q_\alpha (p\cdot q)\right) n_\mu \right. \\ &  \left. - \frac{v_2^R v_1^R}{M^2} \epsilon_ {\alpha\mu\nu\rho} p^\nu n^\rho (p\cdot q) + \frac{v_2^R v_1^L}{M^2} \epsilon_ {\alpha\mu\nu\rho} q^\nu n^\rho (p\cdot q) + \frac{v_2^R v_2^R}{M^2} \left[p_\alpha \left(q_\mu (n\cdot p) - p_\mu (n\cdot q)\right) - \eta_{\alpha\mu} \left((n\cdot p)(p\cdot q) - (n\cdot q) p^2\right)\right] \right. \\ &  \left. - \frac{v_2^Rv_2^L}{M^2} \left[q_\alpha \left(p_\mu (n\cdot q) - q_\mu (n\cdot p)\right) - \eta_{\alpha\mu} \left((n\cdot q)(p\cdot q) - (n\cdot p) q^2\right)\right] + \frac{v_2^R v_2^R}{M^2} p_\alpha q_\mu (n\cdot p) \right. \\ &  \left. - \frac{(v_1^R v_1^R - v_2^R v_2^R)}{2 M^2} \left(q_\alpha n_\mu + \eta_{\alpha\mu} (n\cdot q)\right) p^2 + \frac{(v_1^R v_1^R- v_2^R v_2^R - v_5^R)}{M^2} \left(p_\alpha n_\mu + \eta_{\alpha\mu} (n\cdot p)\right) (p\cdot q) \right. \\ &  \left. + \frac{v_3^R}{M^2} \left(p_\alpha (n\cdot q) + q_\alpha (n\cdot p)\right) q_\mu - \frac{v_3^R}{M^2} \left(p_\alpha n_\mu + \eta_{\alpha\mu} (n\cdot p)\right) q^2 + \frac{v_4^R}{M^2} 2 q_\alpha p_\mu (n\cdot q) - \frac{v_4^R}{M^2} \left(q_\alpha n_\mu + \eta_{\alpha\mu} (n\cdot q)\right) (p\cdot q) \right. \\ &  \left. + \frac{v_5^R}{M^2} \left(p_\alpha (n\cdot q) + q_\alpha (n\cdot p)\right) p_\mu + \frac{v_6^R}{M^2} \left[q_\alpha \epsilon_{\mu\nu\rho\sigma} q^\nu p^\rho n^\sigma - \epsilon_{\alpha\mu\nu\rho} q^\nu p^\rho (n\cdot q)\right] + \frac{v_7^R}{M^2} \left[p_\alpha \epsilon_{\mu\nu\rho\sigma} q^\nu p^\rho n^\sigma - \epsilon_{\alpha\mu\nu\rho} q^\nu p^\rho (n\cdot p)\right]\right] \,.
\end{split}
\label{q'2}
\ee

We see that the coefficients of the modified Lorentz transformations depend on the 14 parameters $(v_i^L;v_i^R), i=1,\ldots 7$, appearing in the change of variables, and not on the 4 parameters $c_i$ of the covariant composition law, so that, as we stated in the last point of Sec.~\ref{sec:specific}, at second order the modified Lorentz transformations do not determine a generic nonlinear composition law. 

If we choose again $n_\mu=(1, 0, 0, 0)$ in Eqs.~\eqref{p'2} and~\eqref{q'2} we get
\be
\begin{split}
p_{0}^{\prime}&=p_{0}+\vec{p}\cdot \vec{\xi}-\frac{v_{1}^{L}}{M}q_{0}\left(\vec{p}\cdot \vec{\xi}\right)+\frac{v_{2}^{L}}{M}\left(\vec{p}\wedge\vec{q}\right)\cdot \vec{\xi}+\frac{v_{1}^{L}v_{1}^{L}-v_{2}^{L}v_{2}^{L}-2v_{5}^{L}}{2M^{2}}q_{0}^{2}\left(\vec{p}\cdot \vec{\xi}\right)+\frac{v_{1}^{L}v_{1}^{L}+v_{2}^{L}v_{2}^{L}}{2M^{2}}\vec{q}^{2}\left(\vec{p}\cdot \vec{\xi}\right) \\
& +\frac{v_{1}^{L}v_{1}^{R}-v_{3}^{L}}{M^{2}}\vec{p}^{2}\left(\vec{q}\cdot \vec{\xi}\right) +
\frac{v_{1}^{L}v_{1}^{R}-v_{3}^{L}-v_{4}^{L}}{M^{2}}p_{0}q_{0}\left(\vec{p}\cdot \vec{\xi}\right) -
\frac{v_{1}^{L}v_{1}^{R}+v_{4}^{L}}{M^{2}}\left(\vec{p}\cdot \vec{q}\right)\left(\vec{p}\cdot \vec{\xi}\right) \\
& -
\frac{v_{2}^{L}v_{2}^{L}+v_{5}^{L}}{M^{2}}\left(\vec{p}\cdot\vec{q}\right)\left(\vec{q}\cdot\vec{\xi}\right)+\frac{v_{1}^{L}v_{2}^{R}+v_{6}^{L}}{M^{2}}p_{0}\left(\vec{p}\wedge\vec{q}\right)\vec{\xi}+\frac{-v_{1}^{L}v_{2}^{L}+v_{7}^{L}}{M^{2}}q_{0}\left(\vec{p}\wedge\vec{q}\right)\vec{\xi} \, ,
\end{split}
\label{generaltr1}
\ee
\be
\begin{split}
p_{i}^{\prime}&=p_{i}+p_{0}\xi_{i}-\frac{v_{1}^{L}}{M}\left[q_{i}\left(\vec{p}\cdot \vec{\xi}\right)+\left(p\cdot q\right)\xi_{i}\right]-\frac{v_{2}^{L}}{M^{2}}\left(q_{0}\epsilon_{ijk}p_{j}\xi_{k}-p_{0}\epsilon_{ijk}q_{j}\xi_{k}\right)+\frac{v_{1}^{L}v_{1}^{R}-v_{3}^{L}-v_{4}^{L}}{M^{2}}p_{0}^{2}q_{0}\xi_{i} \\
&+
\frac{v_{1}^{L}v_{1}^{L}-v_{2}^{L}v_{2}^{L}-2v_{5}^{L}}{2M^{2}}p_{0}q_{0}^{2}\xi_{i}+\frac{-v_{1}^{L}v_{1}^{R}-v_{2}^{L}v_{2}^{R}+v_{4}^{L}}{M^{2}}\left(\vec{p}\cdot \vec{q}\right)p_{0}\xi_{i}+\frac{-v_{1}^{L}v_{1}^{L}+2v_{2}^{L}v_{2}^{L}+v_{5}^{L}}{M^{2}}\left(\vec{p}\cdot \vec{q}\right)q_{0}\xi_{i} \\
&+\frac{v_{2}^{L}v_{2}^{R}+v_{3}^{L}}{M^{2}}\vec{p}^{2}q_{0}\xi_{i} +
\frac{v_{1}^{L}v_{1}^{L}-3v_{2}^{L}v_{2}^{L}}{2M^{2}}p_{0}\vec{q}^{2}\xi_{i}+\frac{v_{2}^{L}v_{2}^{R}-2v_{4}^{L}}{M^{2}}p_{0}q_{i}\left(\vec{p}\cdot \vec{\xi}\right)-\frac{v_{2}^{L}v_{2}^{R}+v_{3}^{L}}{M^{2}}p_{i}q_{0}\left(\vec{p}\cdot \vec{\xi}\right)
\\ &+\frac{v_{2}^{L}v_{2}^{L}-v_{5}^{L}}{M^{2}}p_{0}q_{i}\left(\vec{q}\cdot \vec{\xi}\right)-\frac{2v_{2}^{L}v_{2}^{L}}{M^{2}}p_{i}q_{0}\left(\vec{q}\cdot \vec{\xi}\right) +
\frac{v_{1}^{L}v_{1}^{R}-v_{3}^{L}}{M^{2}}p_{0}p_{i}\left(\vec{q}\cdot \vec{\xi}\right)+\frac{v_{1}^{L}v_{1}^{L}-v_{5}^{L}}{M^{2}}q_{0}q_{i}\left(\vec{p}\cdot \vec{\xi}\right) \\
&-\frac{v_{1}^{L}v_{2}^{L}}{M^{2}}q_{i}\left(\vec{p}\wedge\vec{q}\right)\vec{\xi}+\frac{v_{1}^{L}v_{2}^{R}+v_{6}^{L}}{M^{2}}p_{0}^{2}\epsilon_{ijk}q_{j}\xi_{k}-\frac{v_{6}^{L}}{M^{2}}\left(\vec{p}\cdot\vec{\xi}\right)\epsilon_{ijk}p_{j}q_{k}+\frac{v_{7}^{L}}{M^{2}}p_{0}q_{0}\epsilon_{ijk}q_{j}\xi_{k} \\
&-
\frac{v_{7}^{L}}{M^{2}}\left(\vec{q}\cdot \vec{\xi}\right)\epsilon_{ijk}p_{j}q_{k}-\frac{v_{7}^{L}}{M^{2}}q_{0}^{2}\epsilon_{ijk}p_{j}\xi_{k}-\frac{v_{1}^{L}v_{2}^{R}+v_{6}^{L}}{M^{2}}p_{0}q_{0}\epsilon_{ijk}p_{j}\xi_{k}+\frac{v_{1}^{L}v_{2}^{L}}{M^{2}}\left(p\cdot q\right)\epsilon_{ijk}q_{j}\xi_{k}-\frac{v_{1}^{R}v_{2}^{L}}{M^{2}}\left(p\cdot q\right)\epsilon_{ijk}p_{j}\xi_{k} \,,
\end{split}
\ee
\be
\begin{split}
q_{0}^{\prime}&=q_{0}+\vec{q}\cdot \vec{\xi}-\frac{v_{1}^{R}}{M}p_{0}\left(\vec{q}\cdot \vec{\xi}\right)+\frac{v_{2}^{R}}{M}\left(\vec{q}\wedge\vec{p}\right)\cdot \vec{\xi}+\frac{v_{1}^{R}v_{1}^{R}-v_{2}^{R}v_{2}^{R}-2v_{5}^{R}}{2M^{2}}p_{0}^{2}\left(\vec{q}\cdot \vec{\xi}\right)+\frac{v_{1}^{R}v_{1}^{R}+v_{2}^{R}v_{2}^{R}}{2M^{2}}\vec{p}^{2}\left(\vec{q}\cdot \vec{\xi}\right) \\
& +\frac{v_{1}^{L}v_{1}^{R}-v_{3}^{R}}{M^{2}}\vec{q}^{2}\left(\vec{p}\cdot \vec{\xi}\right) +
\frac{v_{1}^{L}v_{1}^{R}-v_{3}^{R}-v_{4}^{R}}{M^{2}}q_{0}p_{0}\left(\vec{q}\cdot \vec{\xi}\right) -
\frac{v_{1}^{L}v_{1}^{R}+v_{4}^{R}}{M^{2}}\left(\vec{p}\cdot \vec{q}\right)\left(\vec{q}\cdot \vec{\xi}\right) \\
& -
\frac{v_{2}^{R}v_{2}^{R}+v_{5}^{R}}{M^{2}}\left(\vec{p}\cdot\vec{q}\right)\left(\vec{p}\cdot\vec{\xi}\right)-\frac{v_{1}^{R}v_{2}^{L}+v_{6}^{R}}{M^{2}}q_{0}\left(\vec{p}\wedge\vec{q}\right)\vec{\xi}+\frac{v_{1}^{R}v_{2}^{R}-v_{7}^{R}}{M^{2}}p_{0}\left(\vec{p}\wedge\vec{q}\right)\vec{\xi} \, ,
\end{split}
\ee
\be
\begin{split}
q_{i}^{\prime}&=q_{i}+q_{0}\xi_{i}-\frac{v_{1}^{R}}{M}\left[p_{i}\left(\vec{q}\cdot \vec{\xi}\right)+\left(p\cdot q\right)\xi_{i}\right]-\frac{v_{2}^{R}}{M^{2}}\left(p_{0}\epsilon_{ijk}q_{j}\xi_{k}-q_{0}\epsilon_{ijk}p_{j}\xi_{k}\right)+\frac{v_{1}^{L}v_{1}^{R}-v_{3}^{R}-v_{4}^{R}}{M^{2}}q_{0}^{2}p_{0}\xi_{i} \\
&+
\frac{v_{1}^{R}v_{1}^{R}-v_{2}^{R}v_{2}^{R}-2v_{5}^{R}}{2M^{2}}q_{0}p_{0}^{2}\xi_{i}+\frac{-v_{1}^{L}v_{1}^{R}-v_{2}^{R}v_{2}^{L}+v_{4}^{R}}{M^{2}}\left(\vec{p}\cdot \vec{q}\right)q_{0}\xi_{i}+\frac{-v_{1}^{R}v_{1}^{R}+2v_{2}^{R}v_{2}^{R}+v_{5}^{R}}{M^{2}}\left(\vec{p}\cdot \vec{q}\right)p_{0}\xi_{i} \\
&+\frac{v_{2}^{L}v_{2}^{R}+v_{3}^{R}}{M^{2}}\vec{q}^{2}p_{0}\xi_{i} +
\frac{v_{1}^{R}v_{1}^{R}-3v_{2}^{R}v_{2}^{R}}{2M^{2}}q_{0}\vec{p}^{2}\xi_{i}+\frac{v_{2}^{L}v_{2}^{R}-2v_{4}^{R}}{M^{2}}q_{0}p_{i}\left(\vec{q}\cdot \vec{\xi}\right)-\frac{v_{2}^{L}v_{2}^{R}+v_{3}^{R}}{M^{2}}q_{i}p_{0}\left(\vec{q}\cdot \vec{\xi}\right)
\\ &+\frac{v_{2}^{R}v_{2}^{R}-v_{5}^{R}}{M^{2}}q_{0}p_{i}\left(\vec{p}\cdot \vec{\xi}\right)-\frac{2v_{2}^{R}v_{2}^{R}}{M^{2}}q_{i}p_{0}\left(\vec{p}\cdot \vec{\xi}\right) +
\frac{v_{1}^{L}v_{1}^{R}-v_{3}^{R}}{M^{2}}q_{0}q_{i}\left(\vec{p}\cdot \vec{\xi}\right)+\frac{v_{1}^{R}v_{1}^{R}-v_{5}^{R}}{M^{2}}p_{0}p_{i}\left(\vec{q}\cdot \vec{\xi}\right) \\
&+\frac{v_{1}^{R}v_{2}^{R}}{M^{2}}p_{i}\left(\vec{p}\wedge\vec{q}\right)\vec{\xi}+\frac{v_{1}^{R}v_{2}^{L}+v_{6}^{R}}{M^{2}}q_{0}^{2}\epsilon_{ijk}p_{j}\xi_{k}+\frac{v_{6}^{R}}{M^{2}}\left(\vec{q}\cdot\vec{\xi}\right)\epsilon_{ijk}p_{j}q_{k}+\frac{v_{7}^{R}}{M^{2}}p_{0}q_{0}\epsilon_{ijk}p_{j}\xi_{k} \\
&+
\frac{v_{7}^{R}}{M^{2}}\left(\vec{p}\cdot \vec{\xi}\right)\epsilon_{ijk}p_{j}q_{k}-\frac{v_{7}^{R}}{M^{2}}p_{0}^{2}\epsilon_{ijk}q_{j}\xi_{k}-\frac{v_{1}^{R}v_{2}^{L}+v_{6}^{R}}{M^{2}}p_{0}q_{0}\epsilon_{ijk}q_{j}\xi_{k}+\frac{v_{1}^{R}v_{2}^{R}}{M^{2}}\left(p\cdot q\right)\epsilon_{ijk}p_{j}\xi_{k}-\frac{v_{1}^{L}v_{2}^{R}}{M^{2}}\left(p\cdot q\right)\epsilon_{ijk}q_{j}\xi_{k} \,.
\end{split}
\label{generaltr4}
\ee
These are the expressions that generalize Eq.~\eqref{MLT-1st} to order $(1/M)^2$: they are non-linear modifications of the Lorentz transformations to second order in the two-particle system that make $p^2$ and $q^2$ invariant, and their coefficients depend on 14 parameters that can be identified as the parameters of a generic change of variables in the two-particle system. As remarked before, they can be extended by a change of basis to make modified second-order dispersion relations invariant. The calculations including a change of basis are rather involved, but in the following subsection we will consider an important case for which they are greatly simplified. It is the case in which the corrections to SR start directly at second order. 

\subsection{Change of variables and change of basis starting at second order}
\label{sec:second order}

As we saw in the Introduction, a few phenomenological indications seem to suggest that the possible corrections in a modified kinematics should start at second order. In fact there exist also some theoretical arguments favoring second-order over first-order corrections, such as thought-experiments involving the Heisenberg microscope or black holes, which conclude quite generally that quantum gravity generates $G\propto m_P^{-2}$ corrections to, for example, the standard Heisenberg uncertainty principle~\cite{Garay1995,Hossenfelder:2012jw}, or the fact that, from the point of view of effective field theories, 
and in the context of supersymmetry, $d=6$ Lorentz-violating operators (corresponding to $M^{-2}$ corrections) are able to suppress unwanted Lorentz violations at low energies which are generated through radiative corrections, while they are however unavoidable if $d=5$ Lorentz-violating operators ($M^{-1}$ corrections to SR) are present~\cite{Collins2004,Bolokhov2005,Kislat2015}.

In this subsection, then, we consider the relevant case in which there are no first-order corrections to SR, and we will study the most general modification of SR at second order in the same way as we did in Sec.~\ref{sec:covariant} for the first-order case.

The expressions corresponding to the modified Lorentz transformation and composition law in the case of a change of variables are obtained easily from Eqs.~\eqref{p'2},\eqref{q'2},\eqref{cl2} by making $v_1^L, v_1^R , v_2^L, v_2^R$ equal to zero. 

If corrections to SR start at second order it is straightforward to make a change of basis at second order
\be
X_\mu = \hat{X}_\mu + \frac{b_4}{M^2} n_\mu \hat{X}^2 (n\cdot\hat{X}) + \frac{b_5}{M^2} \hat{X}_\mu (n\cdot\hat{X})^2 + \frac{b_6}{M^2} n_\mu (n\cdot\hat{X})^3 ,\
\label{eq:basiscov2}
\ee
leading to a dispersion relation
\be
m^2 = p^2 = \hat{p}^2 + \frac{2(b_4+b_5)}{M^2} \hat{p}^2 (n\cdot\hat{p})^2 + \frac{2 b_6}{M^2} (n\cdot\hat{p})^4 .\
\label{drhat2}
\ee

If we choose $n_\mu=(1, 0, 0, 0)$ in Eq.~\eqref{drhat2}, we get 
\be
m^{2}=\hat{p}_{0}^{2}-{\vec{\hat{p}}}^{2}+\frac{\alpha_{3}}{M^2}\left(\hat{p}_{0}\right)^{4}+\frac{\alpha_{4}}{M^2}(\hat{p}_{0})^{2}\vec{\hat{p}}^{2} \,
\ee
which is the modified dispersion relation that generalizes Eq.~\eqref{eq:MDR} to the case of corrections to SR starting at second order, and where
\be
\alpha_{3}=2(b_{4}+b_{5}+b_{6})\qquad\alpha_{4}=-2(b_{4}+b_{5}).
\ee
The change of basis modifies the composition law so that the second-order coefficients in Eqs.~\eqref{MCLpar-ch-2nd} become:
\begin{align}
\beta_3 &= c_1 - v_3^R -b_4 & \beta_4 &= c_2  -  v_3^L -b_4 & \beta_5 &=  c_3 -  v_4^L -  v_5^R-2b_4  \nonumber \\
\beta_6 &=  c_4 -  v_5^L -  v_4^R -2 b_4 & \beta_7 &=  v_3^L+ v_4^L +  v_5^R- 3b_5- 3b_6 & \beta_8 & =  v_3^R +  v_5^L + v_4^R -3b_5 - 3b_6 \nonumber \\
\gamma_4 &= c_1 & \gamma_5 &= c_2  & \gamma_6 &=c_3 \nonumber \\ \gamma_7 &=c_4 & \gamma_8 &= v_4^L-b_5 & \gamma_9 &= v_4^R-b_5 \nonumber \\
\gamma_{10} & =v_3^L+ v_5^R-2b_5 &
\gamma_{11} &=v_3^R + v_5^L-2b_5 & \gamma_{12} &=v_6^L-v_7^R \nonumber \\ \gamma_{13} &=v_7^L-v_6^R\,.
\label{MCLpar-ch-only-2nd}
\end{align}
From these expressions we can get the golden rules (the relations between the coefficients of the modified dispersion relation and the modified composition law) at second order:
\be
\begin{split}
\beta_3 + \beta_6 - \gamma_4 -\gamma_7 + \gamma_9 +\gamma_{11} = 
\beta_4 + \beta_5 - \gamma_5 -\gamma_6 + \gamma_8 +\gamma_{10} &= \frac{3}{2} \, \alpha_4 \\
\beta_7 - \gamma_8 -\gamma_{10}=\beta_8 - \gamma_9 -\gamma_{11} &= -\frac{3}{2} \, (\alpha_3 + \alpha_4).
\end{split}
\label{gr-at-2nd}
\ee

\subsection{Generalized kinematics and the choice of momentum variables}
\label{sec:choice}

In the previous subsections we have used the change of variables as a mathematical tool to identify the non-linear Lorentz transformations to second order in $(1/M)$ that leave invariant the set of dispersion relations in the two-particle system. As we commented in the Introduction, there has been a debate on the literature about the physical meaning of the choice of momentum variables in extensions of SR. In the absence of a dynamical theory, it is difficult to give a definite answer on this question, although, as we also explained above, from the algebraic or geometric points of view a change of variables is qualitatively different from a change of basis, which in those contexts is mathematically irrelevant.

Nevertheless one could ask whether every generalized relativistic kinematics can be seen as a consequence of a change in the (possibly non-trivial, or physically inequivalent) assignation of momentum variables over the kinematics of SR. In Sec.~\ref{sec:firstorder} we showed that this was indeed the case at first order. We will see now that this is not completely so in the case of the generalized kinematics at second order.

As we have seen, at second order the non-linearity of the Lorentz transformations and the correspondent non-linearity in the composition law (the terms proportional to the parameters of the change of variables $v_i^L, v_i^R$), is due to a choice of momentum variables. We had however another relevant ingredient in this case:  
the covariant composition law~\eqref{ccl2}. We will see now that it is not possible to generate an arbitrary covariant composition law by a change of variables and a change of basis.

Let us start with a linear (additive) composition law in the variables $\left\{ \hat{P}\,,\hat{Q}\right\}$, and make from here a covariant change of basis
\be
\tilde{P}_{\mu}=\hat{P}_{\mu}\left(1+\frac{b}{M^{2}}\hat{P}^{2}\right)
\ee  
that leaves invariant the dispersion relation because $\hat{P}^{2}$ is an invariant constant. Following the procedure indicated by Eq.~\eqref{eq:MCLdef} we obtain 
\be
\left[\tilde{P}\tildebigo \tilde{Q}\right]_{\mu}=\tilde{P}_{\mu}+\tilde{Q}_{\mu}-\frac{b}{M^{2}}\tilde{P}_{\mu}\tilde{Q}^{2}-\frac{b}{M^{2}}\tilde{Q}_{\mu}\tilde{P}^{2}-\frac{2b}{M^{2}}\tilde{P}_{\mu}(\tilde{P}\cdot\tilde{Q})-\frac{2b}{M^{2}}\tilde{Q}_{\mu}(\tilde{P}\cdot\tilde{Q}) \,.
\ee
Now we make a change of variables that does not mix momenta in the dispersion relation:
\be
\tilde{P}_{\mu}=P_{\mu}+\frac{v^{L}}{M^{2}}\left(Q_{\mu}P^{2}-P_{\mu}(P\cdot Q)\right)\qquad\tilde{Q}_{\mu}\,=\,Q_{\mu}+\frac{v^{R}}{M^{2}}\left(P_{\mu}Q^{2}-Q_{\mu}(P\cdot Q)\right) \,.
\ee
In this way we finally obtain a covariant composition law of the form
\begin{align}
\left[P\bigo Q\right]_{\mu}=P_{\mu}+Q_{\mu}+\frac{v^{R}-b}{M^{2}}P_{\mu}Q^{2}+\frac{v^{L}-b}{M^{2}}Q_{\mu}P^{2}-\frac{v^{L}+2b}{M^{2}}P_{\mu}(P\cdot Q)-\frac{v^{R}+2b}{M^{2}}Q_{\mu}(P\cdot Q) \,.
\label{cclcv}
\end{align}
If we compare Eq.~\eqref{cclcv} with Eq.~\eqref{ccl2}, we see that there are four parameters in Eq.~\eqref{ccl2}, while the composition law~\eqref{cclcv} contains only three. This means that we can not obtain the most general covariant composition law through a change of variables and a change of basis.
 
To sum up, we see that 17 out of the 18 parameters $(v_i^L, v_i^R , c_i)$ correspond to a choice of momentum variables, leaving out a combination of the $c_i$ that do not correspond to a change of variables or a change of basis. Independently of the physical meaning of a change of variables, this result shows that not every generalization of the relativistic kinematics of SR can be reduced to a choice of momentum variables.

\section{Relation with the formalism of Hopf algebras}
\label{sec:Hopf}

Having obtained the expressions for a generic modified kinematics up to second order in a power expansion of $(1/M)$, we will compare our results with the most studied extension of the Poincaré algebra, which is the Hopf algebra of $\kappa$-Poincaré. This is a single-parameter [$\kappa$, our $(1/M)$] deformation of the commutation relations of the $(P_0,P_i,J_i,N_i)$ generators of the Poincaré algebra, which has a structure of co-algebra, and has therefore a co-product operation that can be put in correspondence with the composition law or the deformed boosts of DSR in the appropriate basis of the algebra~\cite{Kowalski-Glikman2002}. Since our expressions of Sec.~\ref{sec:second} correspond to an extension of SR for which the dispersion relations are those of SR, we will compare them with $\kappa$-Poincaré in the classical basis. One can read from Ref.~\cite{Borowiec2010} the co-product of boosts and momenta in this basis:
\be
\Delta\left(N_{i}\right)=N_{i}\otimes\mathbb{1}+\left(\mathbb{1}-\frac{P_{0}}{M}+\frac{P_{0}^{2}}{2M^{2}}+\frac{\vec{P}^{2}}{2M^{2}}\right)\otimes N_{i}-\frac{1}{M}\epsilon_{ijk}P_{j}\left(\mathbb{1}-\frac{P_{0}}{M}\right)\otimes J_{k}
\label{co-boost}
\ee
\begin{align}
\Delta\left(P_{0}\right)&=P_{0}\otimes\left(\mathbb{1}+\frac{P_{0}}{M}+\frac{P_{0}^{2}}{2M^{2}}-\frac{\vec{P}^{2}}{2M^{2}}\right)+\left(\mathbb{1}-\frac{P_{0}}{M}+\frac{P_{0}^{2}}{2M^{2}}+\frac{\vec{P}^{2}}{2M^{2}}\right)\otimes P_{0}+\frac{1}{M}P_{m}\left(\mathbb{1}-\frac{P_{0}}{M}\right)\otimes P_{m} 
\label{co-p0} \\
\Delta\left(P_{i}\right)&=P_{i}\otimes\left(\mathbb{1}+\frac{P_{0}}{M}+\frac{P_{0}^{2}}{2M^{2}}-\frac{\vec{P}^{2}}{2M^{2}}\right)+\mathbb{1}\otimes P_{i}
\label{co-pi}\, .
\end{align}
From the form of these expressions it is evident that $[\Delta(N_j),C\otimes\mathbb{1}]=[\Delta(N_j),\mathbb{1}\otimes C]=0$, since $C$, the Casimir of the algebra, commutes with all of the $(P_0,P_i,J_i,N_i)$ generators. This means that the Casimir of the algebra is trivially extended to the tensor product of the algebras, or, equivalently, as was described in Sec.~\ref{sec:summary}, that the dispersion relation of every particle in the two-particle system is a function of the momentum of that particle.

To match these algebraic expressions with the kinematical language we have used in this paper, let us consider that the generators of the Poincaré algebra are operators acting in the basis of the momentum operator, $P_\mu \p=p_\mu \p$. The generators of boosts $N_j$ satisfy
\be
 \pp = (\mathbb{1}-i\xi_j N_j+\mathcal{O}(\xi^2)) \p
\ee
where $\pp\equiv \p'$ is the transformed state from $\p$ with a boost. From the previous equation, and working up to order $\mathcal{O}(\xi^2)$,
\be
-i\xi_j [N_j,P_\mu]\p = - i\xi_j (N_j P_\mu-P_\mu N_j)\p = p_\mu (\pp-\p)-p'_\mu\pp+p_\mu\p = (p-p')_\mu\pp=(p-p')_\mu\p+\mathcal{O}(\xi^2) \,.
\label{deriv}
\ee
From here we get
\be
p'_\mu = p_\mu + i\xi_j \fj,
\label{relac-1}
\ee
where $\fj$ are the eigenvalues of $[N_j,P_\mu]$, which is a function of the $P_\mu$, that is, 
\be
f_j(P_\mu)\p\equiv [N_j,P_\mu]\p = \fj \p\,.
\ee

The previous relations can be generalized to the two-particle system. In this case, we define
\be
(P_\mu \otimes \mathbb{1})\pqp = p'_\mu\pqp \quad \quad
(\mathbb{1} \otimes P_\mu)\pqp = q'_\mu\pqp 
\ee
and the generators of co-boosts, $\Delta(N_j)$, satisfy
\be
\pqp = (\mathbb{1}-i\xi_j \Delta(N_j)+\mathcal{O}(\xi^2)) \pq\,.
\ee
Then, Eq.~\eqref{relac-1} is generalized to
\be
p'_\mu = p_\mu + i\xi_j \fja \quad \quad q'_\mu=q_\mu+i\xi_j\fjb\,,
\label{co-transformed}
\ee 
where $\fja$ and $\fjb$ are the eigenvalues of $[\Delta(N_j),P_\mu\otimes \mathbb{1}]$ and $[\Delta(N_j),\mathbb{1}\otimes P_\mu]$, respectively:
\be
[\Delta(N_j),P_\mu\otimes \mathbb{1}] \pq = \fja \pq \quad \quad [\Delta(N_j),\mathbb{1}\otimes P_\mu]\pq = \fjb\pq\,.
\label{co-transformed2}
\ee
Finally, the co-product $\Delta(P_\mu)$ acts in the momentum space of the two-particle system, such that
\be
\Delta(P_\mu)\pq= \cop\pq\,.
\label{coprod-CL}
\ee

Let us now make explicitly the correspondence between our framework and that of $\kappa$-Poincaré. From Eq.~\eqref{coprod-CL} and Eqs.~\eqref{co-p0} and \eqref{co-pi}, the composition law that corresponds to $\kappa$-Poincaré in the classical basis is
\be
\begin{split}
(p\oplus q)_0&=p_0+q_0+\frac{\vec{p}\cdot\vec{q}}{M}+\frac{p_0}{2M^2}\left(q_0^2-\vec{q}^2\right) + \frac{q_0}{2M^2}\left(p_0^2+\vec{p}^2\right) - \frac{p_0}{M^2}(\vec{p}\cdot \vec{q})\\
(p\oplus q)_i&=p_i+q_i+\frac{q_0 p_i}{M} + \frac{p_i}{M^2}\left(q_0^2-\vec{q}^2\right)\,.
\end{split}
\label{eq:kappa-CL}
\ee
On the other hand, from the co-product of the boost, Eq.~\eqref{co-boost}, and using Eqs.~\eqref{co-transformed} and~\eqref{co-transformed2}, together with the standard commutation relations $[N_i,P_0]=-iP_j$, $[N_i,P_j,]=i\delta_{ij}P_0$, $[J_i,P_j]=i\epsilon_{ijk}P_m$ (remember that we are working in the classical basis), we get
\be
\begin{split}
p'_0&=p_0+\vec{p}\cdot\vec{\xi} \quad \quad \quad \quad p'_i=p_i+p_0\xi_i \,, \\
q'_0&=q_{0}+\vec{q}.\vec{\xi}\left(1-\frac{p_{0}}{M}+\frac{p_{0}^{2}}{2M^{2}}+\frac{\vec{p}^{2}}{2M^{2}}\right)
\\
q'_{i}&=q_{i}+q_{0}\xi_{i}\left(1-\frac{p_{0}}{M}+\frac{p_{0}^{2}}{2M^{2}}+\frac{\vec{p}^{2}}{2M^{2}}\right)+(\vec{p}\cdot\vec{q})\xi_{i}\left(\frac{1}{M}-\frac{p_{0}}{M^{2}}\right)+\vec{q}\cdot\vec{\xi}\left(-\frac{p_{i}}{M}+\frac{p_{0}p_{i}}{M^{2}}\right) \,.
\end{split}
\label{eq:kappa-transformed}
\ee
Comparing Eq.~\eqref{eq:kappa-CL} with Eq.~\eqref{generalCL}, and Eq.~\eqref{eq:kappa-transformed} with Eqs.~\eqref{generaltr1}-\eqref{generaltr4}, we see that our framework reproduces $\kappa$-Poincaré in the classical basis with 
\[
v_{1}^{R}=1\qquad c_{1}=c_{3}=\frac{1}{2} \,,
\]
and with the rest of the parameters equal to zero. As expected, $\kappa$-Poincaré is a particular case of our general framework that includes relativistic kinematics beyond SR at 
first and second order in the power expansion of $\kappa$ ($1/M$).
 
\section{Conclusions}
\label{sec:concl}

In this paper we have shown how to obtain second-order (in an expansion on the inverse of a high energy scale $M$) relativistic kinematics beyond SR. We have been able to give a systematic procedure by using a mathematical trick: a change of variables from momentum variables that transform linearly in the two-particle momentum space and a change of basis in the one-particle momentum space. In fact we showed in Sec.~\ref{sec:firstorder} that this procedure produces the most general relativistic kinematics at first order; it allows one to arrive to the same results that were derived in  Ref.~\cite{Carmona2012} by imposing the relativity principle, and gives a simple interpretation of the mathematical relations that the relativity principle imposes on the coefficients of the modified dispersion relation, the modified composition law, and the modified Lorentz transformations: all of these coefficients are expressions that involve the same parameters appearing in the change of variables and the change of basis.

We have also related different approaches to extensions of SR: the algebraic, the geometric, and the DSR perspectives. In doing so, we have made a distinction between two types of relativistic kinematics beyond SR: those which come from a change of variables, and those which are derived by a simple change of basis. This distinction may be of help for trying to answer the non-resolved question on the meaning of the choice of momentum variables and the physical inequivalence between SR and these extensions. For example, from the algebraic and the geometric points of view, a change of basis represents just a choice of coordinates in the algebra or in the geometry of the momentum space, and has therefore no physical meaning. In fact, in Ref.~\cite{Kowalski-Glikman2002} it was suggested that the structure of space-time could be the common ingredient to different bases in the context of $\kappa$-Poincaré; although the co-product is different for every basis, when one introduces space-time in a certain way (through a pairing compatible with the co-product, see Ref.~\cite{Kowalski-Glikman2002}), one arrives to a non-commutative ($\kappa$-Minkowski) space-time that has the same algebra independently of the basis. A generalization of this result to generic second-order relativistic kinematics, including alternative ways of introducing the space-time, will be presented elsewhere~\cite{Carmona:2016b}.

The mathematical procedure discussed above has allowed us to construct relativistic kinematics at second order, generalizing the formulas of modified composition laws and modified Lorentz transformations that were obtained at first order in Ref.~\cite{Carmona2012} to much more involved expressions at second order. Another important point was the introduction of covariant notation, which also helped to simplify calculations. What is more: formulas in covariant notation are valid without the simplification of rotational invariance.

Another key ingredient was the existence of covariant composition laws, which are absent at first order. Assuming a rotationally invariant kinematics, we have seen that up to second order in the classical basis there are 18 parameters, and 17 of them correspond to a choice of momentum variables. There is a linear combination of the parameters of the covariant composition law that cannot be reproduced by a change of variables from the standard variables of SR. We also saw in Sec.~\ref{sec:Hopf} that our model contains the much studied $\kappa$-Poincaré, which in the classical basis and up to second order, can be reproduced with three parameters different from zero. In this case however the model up to second order can be seen as a choice of momentum variables, since it is possible to reproduce the covariant terms in $\kappa$-Poincaré by a covariant change of variables, Eq.~\eqref{cclcv}, with $b= v^L = -v^R /2 =-1/6$.

The existence of covariant composition laws is therefore related with the problem of the choice of momentum variables, and makes the situation of relativistic kinematics at first and second order very different. In the extreme case of complete arbitrariness in the choice of momentum variables, it suggests that the first non-triviality of non-linear extensions of SR appears at second order only. It is tempting to try to relate this idea with the apparent lack of phenomenological effects of quantum gravity at first order in the Planck mass as mentioned in the Introduction; however it is clear that one needs a dynamical frame (a quantum field theory based on these kinematic extensions of SR) in order to reveal the physical content of the momentum variables and go from these speculative ideas to specific phenomenological predictions. This has to be the subject of future work.

\section*{Acknowledgments}
This work is supported by the Spanish MINECO FPA2015-65745-P (MINECO/FEDER) and Spanish DGIID-DGA Grant No. 2015-E24/2. We acknowledge useful conversations with Salvatore Mignemi, Boris Iveti\'c, and Rina \v{S}trajn.

\end{document}